\documentclass[12pt,preprint]{aastex}
\begin{document}

\title{A Photometric System for Detection of Water and Methane Ices on
  Kuiper Belt Objects\footnote{Based on observations obtained at the
    Gemini Observatory, which is operated by the Association of
    Universities for Research in Astronomy, Inc., under a cooperative
    agreement with the NSF on behalf of the Gemini partnership: the
    National Science Foundation (United States), the Science and
    Technology Facilities Council (United Kingdom), the National
    Research Council (Canada), CONICYT (Chile), the Australian
    Research Council (Australia), Minist\'{e}rio da Ci\^{e}ncia e
    Tecnologia (Brazil) and Ministerio de Ciencia, Tecnolog\'{i}a e
    Innovaci\'{o}n Productiva (Argentina).} $^{,}$\footnote{This paper
    includes data gathered with the 6.5 meter Magellan Telescopes
    located at Las Campanas Observatory, Chile.} }

\author{Chadwick A. Trujillo}
\affil{Gemini Observatory, Northern Operations Center, 670 N. A`ohoku
  Place, Hilo, Hawaii 96720}
\email{trujillo@gemini.edu}
\and
\author{Scott S. Sheppard}
\affil{Carnegie Institution of Washington, 5241 Broad Branch Rd. NW,
  Washington, DC 20015}
\email{sheppard@dtm.ciw.edu}
\and
\author{Emily L. Schaller}
\affil{Lunar and Planetary Laboratory, University of Arizona}
\email{els@lpl.arizona.edu}
\author{[Currently Scheduled to Appear in ApJ March 20, 2011, V730]}

\begin{abstract}
We present a new near-infrared photometric system for detection of
water ice and methane ice in the solar system.  The system consists of
two medium-band filters in the $K$-band region of the near-infrared,
which are sensitive to water ice and methane ice, plus continuum
observations in the $J$-band and $Y$-band.  The primary purpose of
this system is to distinguish between three basic types of Kuiper Belt
Objects (KBOs) --- those rich in water ice, those rich in methane ice,
and those with little absorbance.  In this work, we present
proof-of-concept observations of 51 KBOs using our filter system, 21
of which have never been observed in the near-IR spectroscopically.
We show that our custom photometric system is consistent with previous
spectroscopic observations while reducing telescope observing time by
a factor of $\sim 3$.  We use our filters to identify Haumea
collisional family members, which are thought to be collisional
remnants of a much larger body and are characterized by large
fractions of water ice on their surfaces.  We add $\rm 2009\ YE_{7}$
to the Haumea collisional family based on our water ice band
observations ($J-H_{2}O = -1.03 \pm 0.27$) which indicate a high
amount of water ice absorption, our calculated proper orbital
elements, and the neutral optical colors we measured, $V-R$ = $0.38
\pm 0.04$ , which are all consistent with the rest of the Haumea
family.  We identify several objects dynamically similar to Haumea as
being distinct from the Haumea family as they do not have water ice on
their surfaces.  In addition, we find that only the largest KBOs have
methane ice, and we find that Haumea itself has significantly less
water ice absorption than the smaller Haumea family members.  We find
no evidence for other families in the Kuiper Belt.
\end{abstract}


\section{Introduction}
\label{intro}

Detecting ices on small outer solar system bodies is of utmost
importance because the presence of the most volatile ices, such as
methane ice, may indicate primordial surfaces.  The most massive of
the known Kuiper Belt Objects (KBOs) such as Eris and Pluto, cannot
retain surface ices for the age of the solar system even under
moderate heating, temperatures of $\gtrsim 70$ Kelvin
\citep{2007ApJ...659L..61S}.  Thus, an inventory of volatile ices on
the KBOs has the potential to provide a useful metric of historical
heating.  In the absence of planet-wide resurfacing, those KBOs that
originally had ices but passed too close to the sun would have lost
their ices and those which stayed distant could retain their ices.
Tying the presence of surface ices to KBO orbital classes is one of
the few direct observables which can shed light on the principle
dynamical mechanisms involved in shaping the early Kuiper Belt.

Most ices of interest are bland or nearly so at visible wavelengths,
but many show prominent absorption features at near-infrared
wavelengths.  Although many ices show strong features in the
mid-infrared, at these wavelengths remote sensing via ground-based or
space based telescopes is severely hampered by thermal background
radiation, making detection of ices difficult.  Thus, nearly all
surveys of KBOs to date that have positively identified ices have used
near-infrared data to do so.  The only major surface component which
is not detectable in the near-infrared is the so-called ultra-red
matter \citep{2002AJ....123.1039J}, which appears in the visible and
has not been definitively shown to be correlated with near-infrared
colors.

Many researchers have collected near-infrared spectra of the brightest
KBOs.  Of the $\sim 30$ spectra of these bodies, almost all can be
categorized into three types: water ice dominated, methane ice
dominated and neutral.  For example, the five brightest KBOs (plus
Pluto and Charon) can be divided into either water ice dominated
(Charon, Haumea, Orcus and Quaoar;
\cite{1987Natur.329..522B,2000Sci...287..107B,2004Natur.432..731J,2005ApJ...627.1057T,2007ApJ...655.1172T,2009AA...496..547P,2009ApJ...695L...1F})
or methane ice dominated (Pluto, Makemake, Eris;
\cite{1976Sci...194..835C,2007AJ....133..284B,2005ApJ...635L..97B,2007AA...471..331D,2009AJ....137..315M}).
Many of the fainter bodies have been shown to be featureless.  There
are some bodies that fall outside this broad taxonomy and have
combinations of materials such as Triton, which show significant
absorption of combinations of substances \citep{1999Icar..139..159Q}.
Such bodies may not be adequately described by a simple set of
photometric information.  We note however, that our survey goal is not
to replace the detailed information that spectroscopy can yield, but
to be able to identify population-wide trends and KBOs with unusual
surfaces for follow-up spectroscopy.  Although most bodies can be
grossly characterized as either water or methane types, it is still
important to note that many ices remain extremely difficult to detect
in the near-infrared such as carbon monoxide (CO), carbon dioxide
($\mbox{CO}_2$), and nitrogen ($\mbox{N}_2$) due the
characteristically narrow and weak absorption lines of these ices and
the extreme faintness of the KBOs.  Even for Pluto, which is over a
factor 10 brighter than the next brightest KBO, detection of CO,
$\mbox{CO}_2$ and $\mbox{N}_2$ is challenging with modern
instrumentation \citep{2007AJ....133..420O}.  For this reason, tracing
methane ice in the Kuiper Belt is extremely important as it is one of
the few volatiles that is easily detectable and very sensitive to
thermal history.  Water ice acts as an important comparison ice, as it
is not volatile beyond $\sim 5$ AU in the current solar environment,
although like methane it can be destroyed by galactic cosmic rays and
solar UV photons, likely two of the dominant long-term weathering
processes in the Kuiper Belt \citep{2003EM&P...92..261C}.

There have been several photometric programs, with varying levels of
success.  Numerous attempts at $JHK$ photometry have been published
\citep[see review work by][and references
  therein]{2008ssbn.book..181F}, but unfortunately, these bandpasses
are only serendipitously correlated with methane and water ice
transitions and cannot readily discriminate between the two.  Probably
the most successful set of observations to date are those of
\cite{2010AA...511A..72S}, who surveyed Haumea family members in
search of water ice in the $H$-band.  Their use of the medium bandpass
$H_{s}$ filter (1.52 -- 1.63 \micron) which probes the solid-state
water ice region in KBOs allowed the identification of many Haumea
family members using their photometry and references to literature.
Although the $H_{s}$ filter is sensitive to water ice, it is much less
sensitive to methane ice, which has peak absorption between 1.60 --
1.85 \micron\ in the $H$-band.

In this work, we introduce a simple custom-bandpass three filter
system which allows us to discern between KBOs with water ice, methane
ice and relatively bland surfaces.  We have undertaken
proof-of-concept observations which demonstrate that these filters are
indeed consistent with spectroscopic observations yet typically
require $1/3$ of the telescope time of spectroscopic works.  In
addition, we have observed all of the Haumea family candidates
brighter than $V = 22$ from \cite{2007AJ....134.2160R} and we identify
which members have surfaces rich in water ice.

\section{Experimental Design}

\subsection{Custom Filter Bandpasses}
\label{filterdesign}

Our filter set was chosen by simulating synthetic spectra of water
ice, methane ice and neutral KBO surfaces as observed through the
Earth's atmosphere using data from the Gemini/NIRI Integration Time
Calculator to account for instrumental transmission and thermal
emission. For the methane ice filter, regions in the $K$-band were
favored over the $H$-band for 4 reasons: (1) the $H$-band contains
many more strong telluric OH emission lines than the $K$-band, which
are variable on $\sim 5$ minute timescales and may affect photometry,
(2) transitions for methane and water in the $K$-band are deeper than
in the $H$-band and are more widely separated from each other, (3) a
wider bandpass without strong atmospheric extinction is available in
the $K$-band and (4) the Gemini North telescope, used for the faintest
bodies in this program, is optimized for near-infrared observations
and thus has low emissivity and low thermal contamination in the
$K$-band. The possibility of double-bandpass ``notch'' filters
including ice transitions in both the $H$-band and the $K$-band while
excluding the sky between were also considered, but estimated signal
gain was not sufficient to warrant the inclusion of the $H$-band
telluric OH emission and the loss of throughput for the
double-bandpass design.

\begin{figure}[htbp]
\plotone{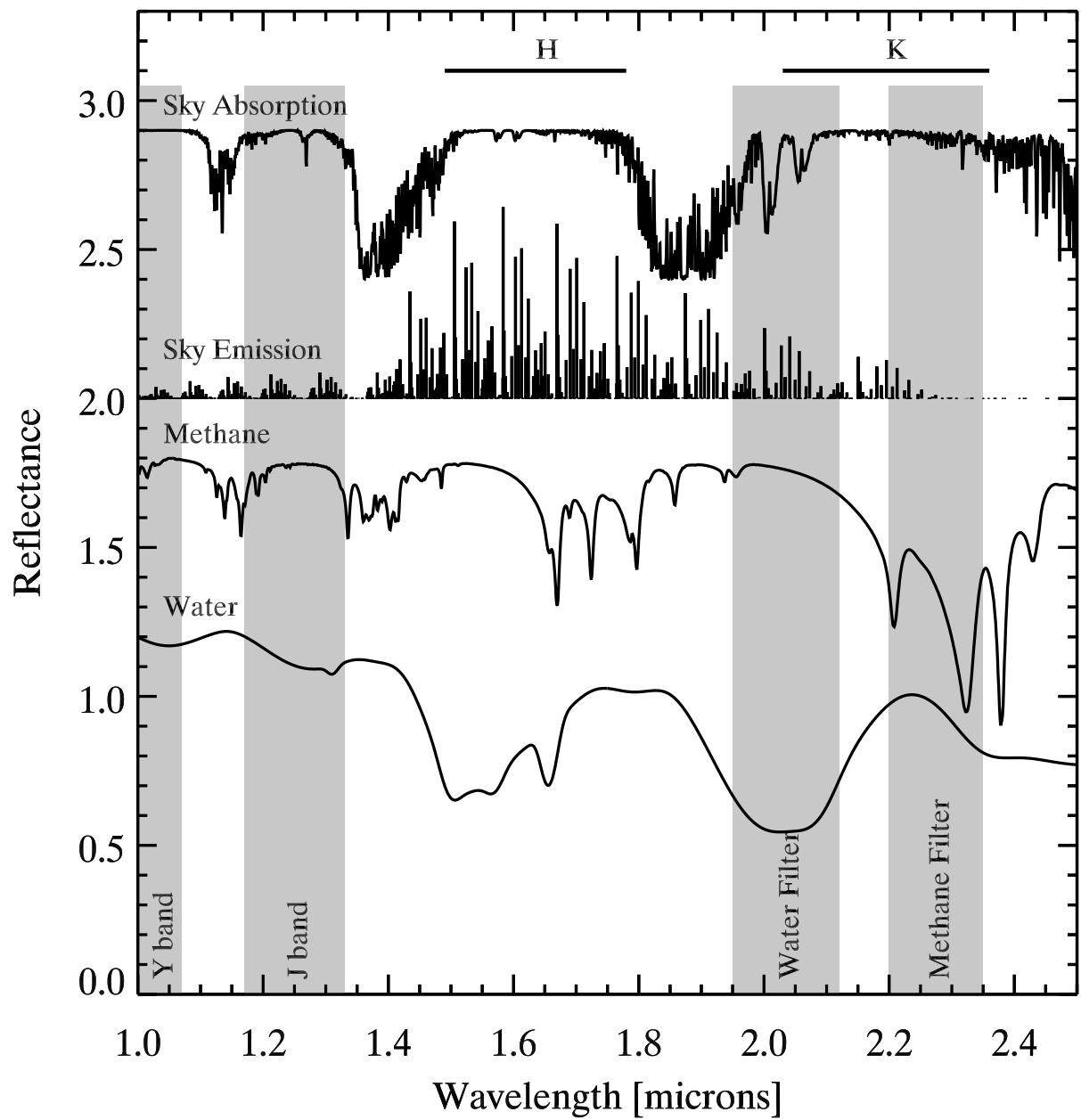}
\vspace*{0.25in} \figcaption{\small Our custom filter bands (grey regions)
  with synthetic water and methane ice spectra overlaid (bottom two
  traces, offset vertically for clarity).  Our primary goal is the
  discrimination between water ice, methane ice, and featureless KBO
  surfaces.  The top two traces show sky absorption and sky OH
  emission, the two main sources of systematic error in the
  near-infrared.  The Y filter and J filter are used to sample the
  continuum by being as wide as possible while still minimizing ice
  and telluric effects.  The ``Water'' filter limits are set by
  telluric absorption on the blue end and the overlapping methane
  absorption on the red end.  The ``Methane'' filter limits are set by
  overlapping water absorption on the blue end and thermal emission on
  the red end.  The methane filter also has some sensitivity to other
  alkanes which have C-H absorption in the methane region.  The top
  portion of the plot shows traditional {\it HK} bandpasses, a poor
  diagnostic for ices.
\label{filters}}
\end{figure}

In order to estimate the amount of absorption seen in the filter,
continuum observations (i.e. observations out of absorption bands) are
also needed as a reference point. Continuum observations must be in
regions that have minimal (1) telluric absorption, (2) atmospheric
emission and (3) ice absorption from the target body. As seen in
Figure~\ref{filters}, the optimum region that satisfies all 3 criteria
is in the $J$-band, where water, methane, and atmospheric
contamination are minimal. There are no regions in either the $H$-band
or $K$-band which are devoid of atmospheric, water and methane ice
absorption. We also collected $Y$-band data on many bodies as a
mitigation against the presence of so called ``ultra-red'' material
\citep{2002AJ....123.1039J}, which can extend throughout the visible
and contaminate the shortest wavelengths of the $J$-band.  After final
analysis, robust results were obtained without the use of the $Y$-band
filter data, but in a few cases, such as Quaoar, which is extremely
red in the visible, it was needed (see Section~\ref{ultrared}).
Final filter bandpasses are listed in Table~\ref{filterdetails}.

\subsection{Filter Manufacturing and Installation}

\begin{figure}[htbp]
\plotone{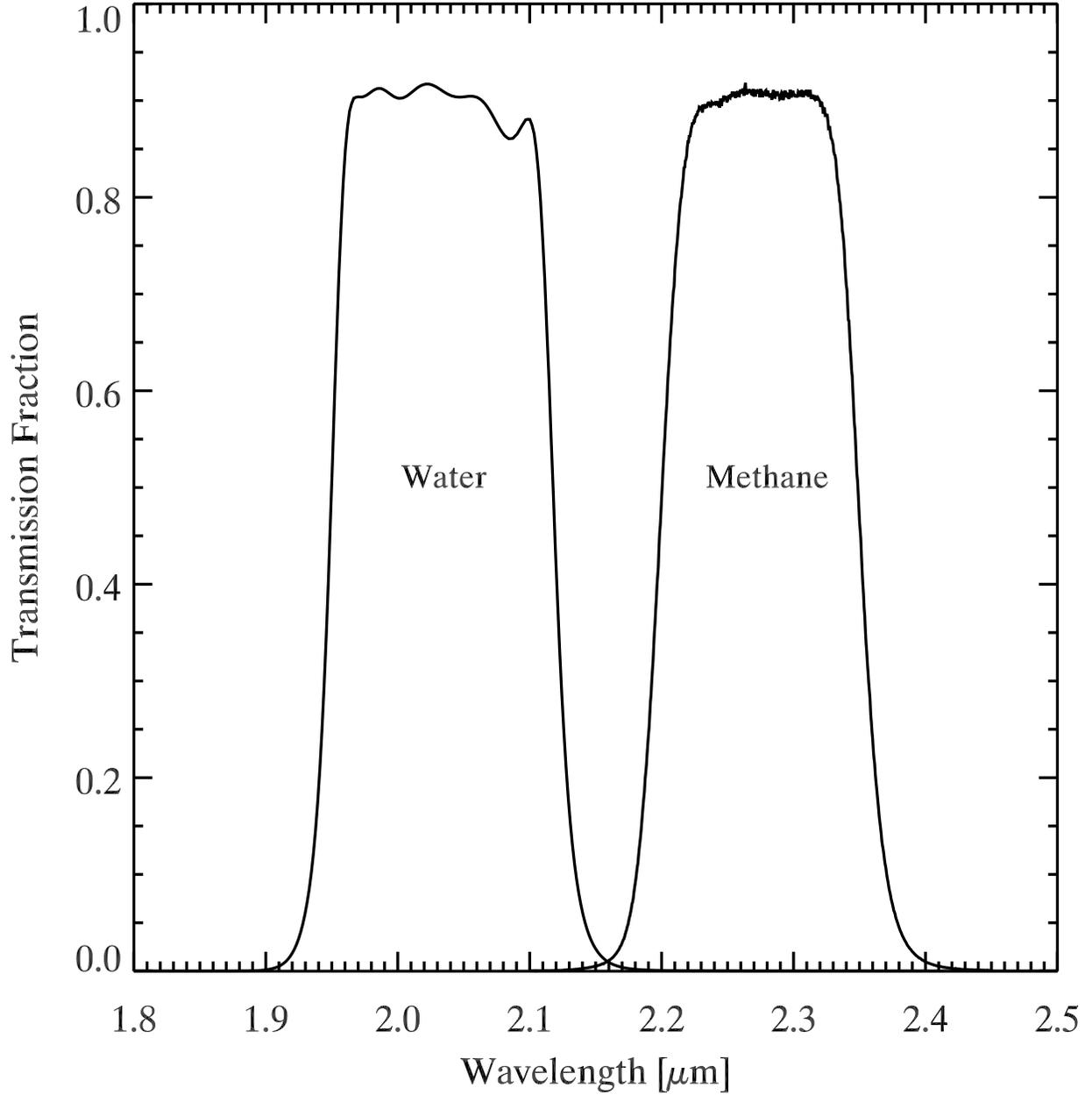}
\vspace*{0.5in} \figcaption{Efficiency curves for our two custom
  filters as a function of wavelength for the water ice filter (left)
  and the methane ice filter (right).
\label{filters-barr}}
\end{figure}

We installed our filters into two telescopes --- The 8.1 m mirror
diameter Frederick C. Gillett Gemini telescope located atop Mauna Kea,
Hawaii and the 6.5 m mirror diameter Magellan Baade Telescope, at Las
Campanas, Chile.  Our filters were manufactured by Barr Associates,
Inc. in Massachusetts, USA, who implemented the bandpasses to our
optical specifications.  Filters with identical bandpasses were
manufactured in the same coating batch to ensure consistent quality
--- only the geometric shape of the filter substrate differed between
Gemini (60 mm diameter, round, 5 mm thick) and Magellan (32mm across,
square, 4 mm thick).  The filter efficiencies are presented in Figure
\ref{filters-barr}.  Filters were installed in late 2008 in the Gemini
North Near InfraRed Imager and Spectrometer \citep[NIRI,
][]{2003PASP..115.1388H} and in the Magellan Baade Persson's Auxiliary
Nasmyth Infrared Camera \citep[PANIC, ][]{2004SPIE.5492.1653M} dewars.
Both Gemini/NIRI and Magellan/PANIC are cryogenic instruments, so
filter installations are performed very rarely, due to the minimum of
2 weeks warm up and cool down cycle time required to service the
instruments.  An extra set of Gemini/NIRI filters was manufactured to
provide contingency in case unexpected manufacturing defects or
installation problems were encountered.

\subsection{Filter Commissioning and Transformation to Common Bandpasses}

\begin{figure}[htbp]
\plotone{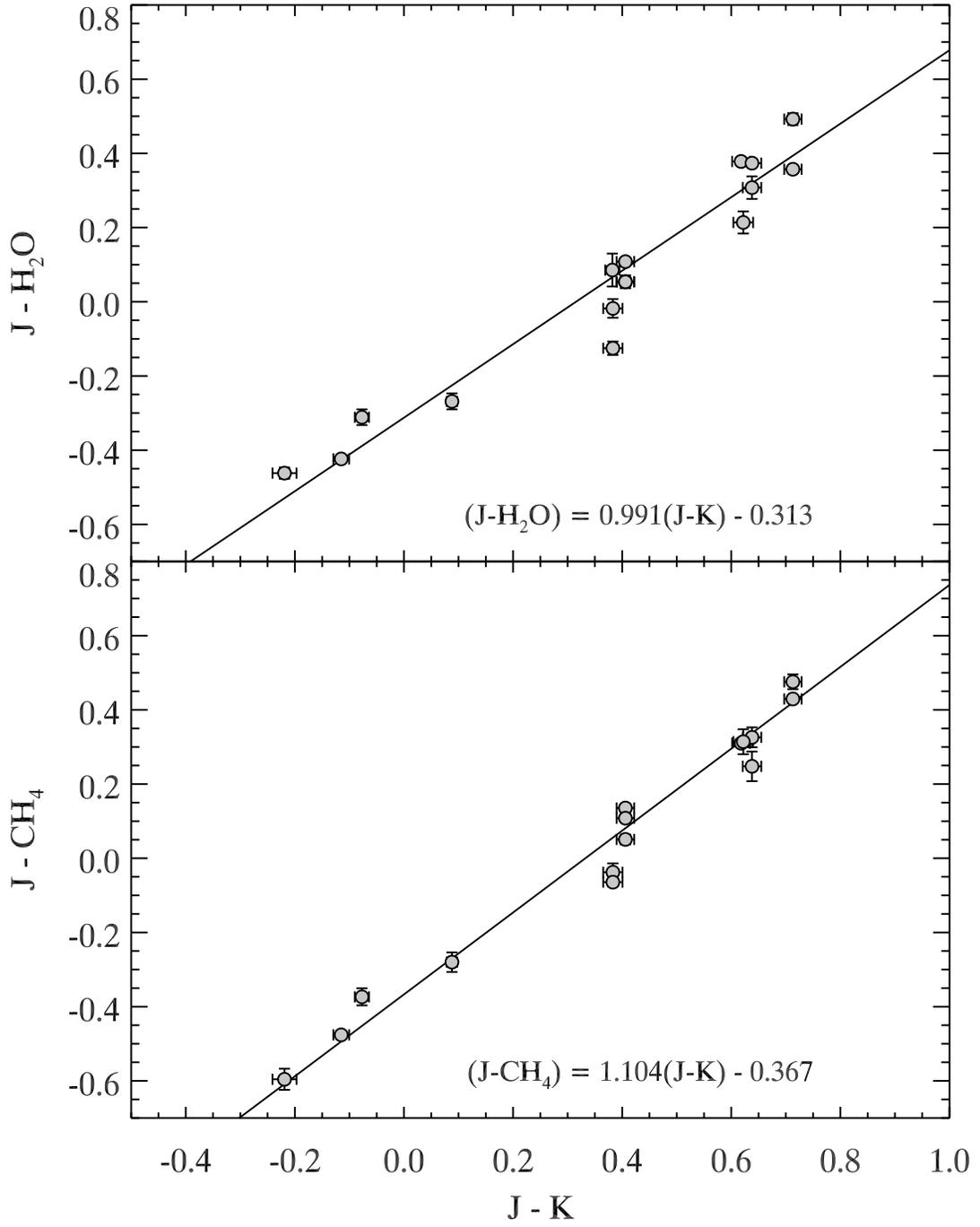}
\vspace*{0.5in} \figcaption{Photometric color calibration for
  Gemini/NIRI.
\label{niristandards}}
\end{figure}

On-sky tests of the filters were performed shortly after installation
to ensure acceptable quality of the filters.  These tests involved
imaging a UKIRT JHK standard star to estimate throughput and zeropoint
and to ensure that image quality was consistent with other standard
filters.  A full characterization with additional targets has since
been performed with a large number of UKIRT standards observed over
several semesters at Gemini North.  We present our custom color
metrics versus standard color metrics in
Equations~\ref{watercolor}~and~\ref{methanecolor}, which are derived
from the data displayed in Figure~\ref{niristandards}.

\begin{equation}
\label{watercolor}
J - H_{2}O = 0.991 (J-K) - 0.313
\end{equation}
\begin{equation}
\label{methanecolor}
J - CH_{4} = 1.104 (J-K) - 0.367
\end{equation}

It is important to note that while these transformations are
applicable to UKIRT standard stars which generally have few
transitions in the $K$-band, they are not applicable to KBOs or other
bodies with icy surfaces, as the entire purpose of our experiment is
to probe the absorption regions of KBO ices.

Since the Gemini North NIRI instrument is also a spectrometer, spectra
were collected of the filters at cryogenic temperatures indicating
that the overall throughput was consistent with that measured in the
Barr laboratory (Figure~\ref{filters-barr}) --- thus no degradation of
filter performance occurred during installation or cooling to
cryogenic temperatures.  This was not possible to do with Magellan
PANIC since it is an imager only, however, the Magellan filters were
manufactured using the same techniques and in the same batch as the
NIRI filters, so we expect that the filters are identical in their
physical properties.  Results of a few science targets observed with
both instruments showed consistency.

\section{Near-Infrared Observational Program}

Near-infrared observations of the brightest targets were performed at
Magellan on four nights spanning two observing runs on UT dates 18--19
October 2008 and 03--04 April 2009.  Weather conditions were
photometric for both runs with typical delivered image quality through
our custom filters of 0.3 -- 0.5 arcsec.  Gemini observations were
conducted in the Gemini North Queue under program IDs GN-2008A-Q-4,
GN-2008B-Q-40, GN-2009A-Q-6, and GN-2009B-Q-30.  Observing conditions
were set at 70 percentile or better for image quality (i.e. $K$-band
image quality of 0.55 arcsec or better at zenith) and 50 percentile
for cloud cover (i.e. photometric).  This typically led to image
quality similar to that seen at Magellan, from 0.3 -- 0.5 arcsec
delivered on the detector.  A more complete summary of the objects and
observational circumstances of the observations appears in
Table~\ref{observations}.

At Magellan, the telescope was tracked at the sidereal rate, while at
Gemini, the telescope was tracked at the non-sidereal rate of the
object.  These two methods were used for ease of execution at the
telescope --- at Magellan any non-sidereal tracking rates must be
updated upon slew by hand while at Gemini the telescope can
automatically compute moving object position and tracking rates from
orbital elements.  Since we were limited to $\sim 60$ second exposures
or shorter due to the thermal sky background, our objects, which
typically have $\sim 3$ arcsec / hour apparent motion, moved less than
$1/5$ of a seeing disk during each image, resulting in image quality
being identical for both the object and background stars no matter
which guiding technique was employed.

While observing a science target, filters were cycled with repeated
visits to the $J$-band to allow for identification and correction of
any lightcurve effects.  A typical observational sequence for a faint
target at Gemini was 3 images taken with 30 second exposure time (3 x
30s) $J$, 6 x 60s $\rm H_{2}O$, 6 x 60s $\rm CH_{4}$ repeated for 1.25
hours while offsetting the telescope between each exposure.  In
addition, a short sequence of 3 x 30s $Y$ was included in the
beginning of the sequence as an additional constraint on any large
scale color trends that might be visible in the $1 \micron$ wavelength
region.

Primary color calibration was done by observing a G2V solar analog
star immediately before or after every science target at an airmass
similar to the science observations.  The aim of collecting these data
was to estimate any telluric extinction or emission in the custom
filter regions, which was particularly important in the $\rm H_{2}O$
filter which contains a known region of atmospheric attenuation, which
could result in a $\sim 3\%$ photometric variation ($\sim 0.03$
magnitudes) if left uncorrected.  These techniques are similar to what
is routinely done for spectroscopic telluric correction to provide
valid spectral information in the $2 \micron$ wavelength region.  This
provides for the simplest possible measure of object reflectance ---
science target colors are compared to the G2V analog taken at similar
airmass to immediately and directly calculate reflectance.

In addition to the G2V solar analog, a UKIRT standard star was
observed every night science observations were conducted.  This
provided information about any possible long term degradation of the
filters which could affect our results if it occurred.  No photometric
change in filter sensitivity of color was observed over the course of
our observations.

\section{Data Reduction}

Basic data reduction was performed in a standard manner for
near-infrared imaging and in a similar manner at both telescopes.
First, daytime flat fields were examined to identify bad pixels
(i.e. those pixels that did not respond linearly with a change of flat
field level).  These pixels were replaced with a median of nearby
pixels on all science and calibration images.  Flat fields at Gemini
were constructed from daytime observations of a calibration source,
while at Magellan sky frames were used.  The science fields were
divided by these flats to correct for the pixel-to-pixel variations
seen in the detector.  At Magellan, the pixel-to-pixel variations were
very minor as the detector was generally cosmetically pristine for a
near-infrared array.  Two deep images were then created for each
target, consisting of all science frames taken of a particular object
(1) digitally shifted to track the object at the non-sidereal rate of
the object of interest, and (2) digitally shifted to track at the
sidereal rate to create a deep image of the star field.  This allowed
for the rejection of science observations where the science object
encountered background stars during the course of the observations.
Finally, synthetic aperture photometry was performed on all of the
science and calibration targets.

Photometric calibration was performed on a nightly basis by
identifying all calibration stars taken on a given night and plotting
the stellar brightness as a function of airmass.  Any deviant
calibration targets were rejected as being stars misidentified as G2V
in the Simbad Astronomical Database \citep{2000AAS..143....9W}, with
spectral typing derived from the Michigan Catalogues of
Two-Dimensional Spectral Types for HD stars
\citep{1975mcts.book.....H}.  Remaining stars were used to create a
telluric correction curve as a function of airmass.  In general, these
observations remained consistent on a nightly basis.  These
photometric calibrations were applied to the science observations to
produce a relative reflectance measurement for each object.

\section{Results}

\begin{figure}[htbp]
\plotone{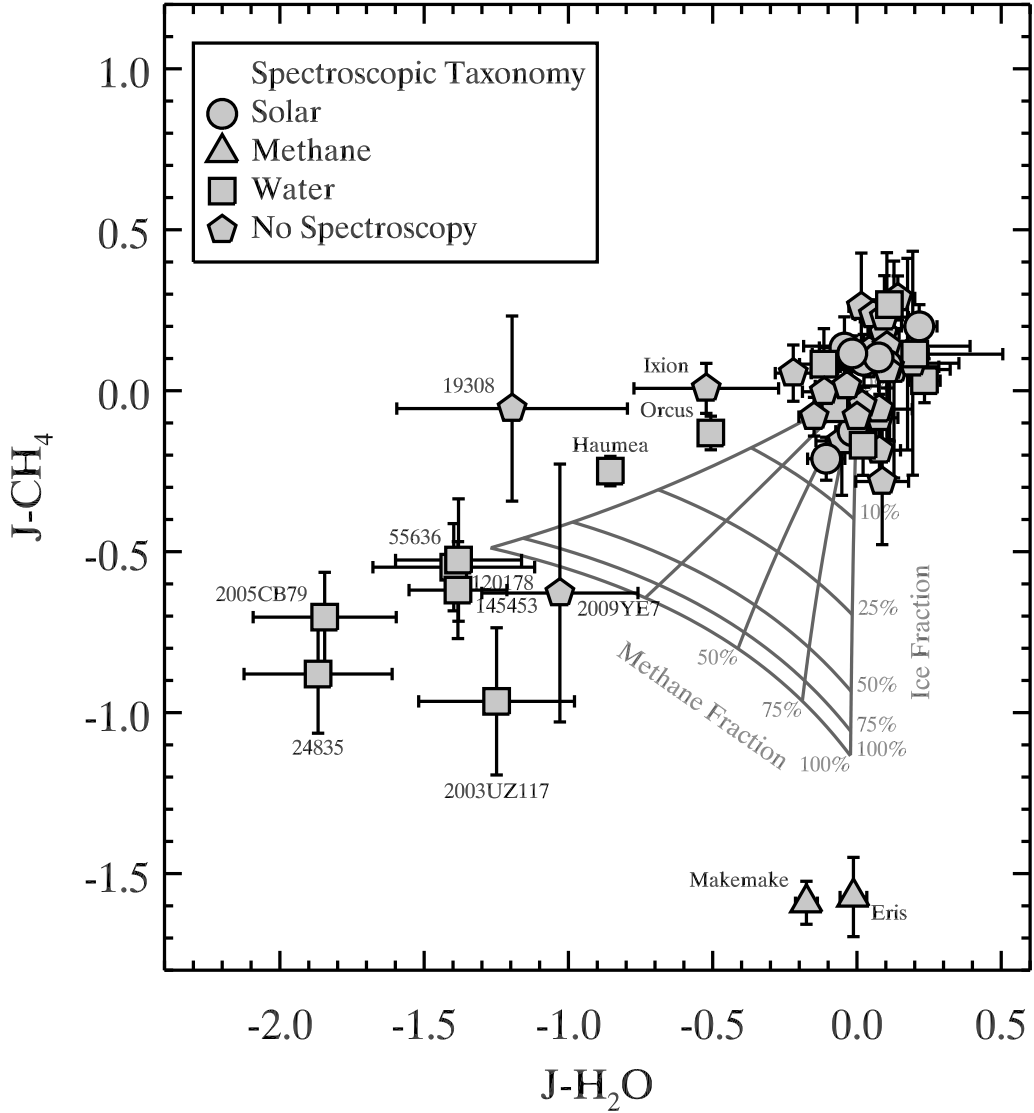}
\figcaption{\small Combined photometry for all objects.  Data
  symbol position is computed from our photometric survey.  Data
  symbol shape is determined from published spectroscopy (see
  Table~\ref{resultstable} for complete list of references).  Objects
  clearly fall into three groups: Methane (Eris and Makemake), Water
  (Haumea family members now including $\rm 2009\ YE_{7}$, Ixion, and
  Orcus) and Neutral / low absorption (all others).  In general there
  is a strong correlation between our photometric work and published
  spectroscopic works for those bodies observed by both.  The
  triangular grey grid represents colors expected for synthetic
  objects with varying percentages of methane, water and total ice
  fraction.  The vertical axis of the grid labelled ``Ice Fraction''
  represents total ice fraction (water and methane combined) with the
  remainder being represented by a neutral absorber.  The diagonal
  axis labelled ``Methane Fraction'' represents the total fraction of
  ice that is methane ice, with water ice being the remainder of the
  ice fraction.  The model details are discussed in full in
  \S\ref{sensitivity}
\label{results}}
\end{figure}

The results of our work are presented in Figure~\ref{results}.  The
basic goal of this work is to demonstrate that our custom filter
bandpass can distinguish between the three basic types of bodies
observed in the Kuiper Belt: water ice, methane ice and neutral
reflectance.  This trend is readily apparent in Figure~\ref{results}
as the vast majority of objects fall into the neutral category in the
upper right.  Most bodies with ice detection have water ice and occupy
the left part of Figure~\ref{results}.  The two known bodies with
methane ice, Makemake and Eris are shown in the bottom right of the
figure, clearly separated from all other objects.

In this proof-of-concept sample, we have studied the brightest KBOs
for two reasons: (1) these bodies require less telescope time to study
and (2) many of these bodies have spectroscopy already collected by
other researchers which provides corroborative evidence that our
survey is performing as expected.  We believe that the current sample
adequately demonstrates that our custom photometric filter system does
indeed distinguish between these basic types of bodies as the three
basic surface types are clearly discernible by eye.  In future works,
we will extend our survey to fainter bodies requiring more telescope
time --- bodies that to date have been inaccessible to spectroscopic
studies.  Even in our current sample, about 40\% of our targets have
no known spectra collected, primarily due to their faintness.

Table~\ref{resultstable} provides verification data for this work.  It
lists the color quantities computed for each object and the surface
type based on this color quantity.  In addition, it also lists
references to published works that provide spectroscopic verification
of our basic surface identification.  Discrepancies between the two
methods occur only for bodies with low absorption, as discussed
in\S~\ref{sensitivity}.  All of the bodies with strong water ice
features show excellent correlation with published spectra.  The
identification of water ice among the Haumea family members, including
recently discovered object $\rm 2009\ YE_{7}$, as well as Ixion and
Orcus, are discussed in section \S~\ref{family}.  For the Haumea
family, we find that the faintest bodies systematically have the
strongest transitions, as discussed in \S~\ref{size}.  In all cases,
our photometric study was done in far less time than spectroscopic
identifications, as discussed in further detail in
\S~\ref{advantages}.

\subsection{Family Members}
\label{family}

\subsubsection{Haumea Family}

The Haumea family members are readily apparent by examining
Figure~\ref{results} and cross-correlating icy bodies with orbital
parameters.  We present this cross-correlation in
Table~\ref{familymembers}, where we list all bodies with $J-H_{2}O <
-0.3$ (water ice bodies) or $J-CH_{4} < -1.3$ (methane ice bodies).
All of the Haumea family members are readily identifiable as having
both water ice on their surfaces and having proper orbital elements,
$i \sim 28\degr$, $a \sim 43.5$ AU and $e \sim 0.1$, as presented in
\cite{2007AJ....134.2160R}.  We have observed all bodies brighter than
$V = 22$ listed as possible Haumea family members on dynamical grounds
from \cite{2007AJ....134.2160R} and report physical associations based
on their high percentage of water ice, as summarized in
Table~\ref{haumeaobs}.  These identifications match previous
spectroscopic works on the subject
\citep{2007Natur.446..294B,2008ApJ...684L.107S,2010AA...511A..72S}.
We also report the identification of a possible new family member,
$\rm 2009\ YE_{7}$, based on its unusually high water ice fraction and
proper orbital elements which are similar to other Haumea members.  We
computed the proper orbital elements for $\rm 2009\ YE_{7}$ using an
average of 20 million years of orbital motion computed by the MERCURY
symplectic integrator \citep{1999MNRAS.304..793C}.

\subsubsection{$\rm 2009\ YE_{7}$}

After determining that $\rm 2009\ YE_{7}$ was a Haumea family member,
we proceeded to collect visible photometry of it to determine if it
was consistent with the other Haumea family members which are neutral
to slightly blue in reflectance, indicating not only that large amounts
of water ice are on their surfaces, but that the water ice is in a
largely pristine form.  

Optical observations of $\rm 2009\ YE_{7}$ were obtained at The
Magellan Clay 6.5 meter telescope on UT March 21, 2010.  The LDSS3
camera has a STA0500A $4064\times4064$ Charge-Coupled Device with
$15\micron$ pixels.  The field of view is about 8.3 arcminutes in
diameter with a scale of $0.189$ arcseconds per pixel.  Each image was
reduced using dithered twilight flats and biases in a standard manner.
Images were acquired through Sloan $g'$, $r'$ and $i'$ filters while
the telescope was auto-guiding at sidereal rates using nearby bright
stars.  Exposure times were between 250 and 300 seconds.  Southern
Sloan standard stars were used to photometrically calibrate the data
\citep{2005AAS...20713111S}.  In order to more directly compare our
results with previous works, the Sloan colors were converted to the
Johnson-Morgan-Cousins BVRI color system using transfer equations from
\cite{2002AJ....123.2121S}, as in \cite{2010AJ....139.1394S}.  These
colors are presented in Table~\ref{ye7visible}.

The optical colors of $\rm 2009\ YE_{7}$ have a spectral slope of
$S(i' > g') = 5.3 \pm 3$, as defined in \cite{2008ssbn.book...91D} and
implemented in \cite{2010AJ....139.1394S}.  Such a spectral slope is
consistent to the other Haumea family members, which have a weighted
mean slope of $S = 1.9 \pm 0.3$, as computed from the values presented
in \cite{2010AA...511A..72S}.  As the visible colors of $\rm
2009\ YE_{7}$ are inconsistent with most other KBOs, which have a mean
$S \sim 20$ \citep{2010AJ....139.1394S}, we consider the visible
colors to be confirmation of the near-infrared photometric and
dynamical identification of $\rm 2009\ YE_{7}$ as a Haumea family
member.

\subsubsection{Other Families}

We find two objects in our survey have much more water ice on their
surface than the other non-Haumea family members, namely Orcus and
Ixion.  These two bodies have low levels of water ice compared to the
Haumea family members, as found by our survey and corroborated by
spectroscopic studies which show possible detection of water ice
\citep{2005ApJ...627.1057T,2008AJ....135...55B}.  These two bodies
both occupy the 3:2 mean motion resonance with Neptune and have
inclinations within 1 degree of one another.  Although a shared origin
might be invoked for these bodies, adding in results of previous
spectroscopic works suggests otherwise. Spectroscopic detections of
water ice show no apparent correlation with inclination among the 3:2
resonance, as (47171), (55638), (84922) and (208996) have inclinations
of 8.4, 16.3, 14.8, and 13.6 degrees, respectively
(Table~\ref{resultstable} and references therein).  Thus, at the
present time, we find no evidence for a family in the 3:2 Resonance,
although clearly there are bodies with moderate water ice present.

The 3:2 Neptune resonance is thought to have ``swept'' up many bodies
shortly after the solar system formed due to Neptune's outward
migration \citep{1993Natur.365..819M}.  Thus, even if there were a
collisional family in the 3:2 resonance, there is likely to be many
non-family members within the 3:2 resonance that have no water ice on
their surfaces.  Identifying a family among interlopers would be very
difficult in the 3:2 resonance unless the collision happened very
recently, of order $\sim 100$ kyr, over which time orbital elements
can be stable for some objects.  Much longer resonance habitation or
large collisional velocities can cause profound departure (and
ejection) of bodies, due to the very fine structure in the 3:2
resonance \citep{2009AJ....138..827T,2008ssbn.book..275M}.

It has been suggested that KBOs may be devoid or somewhat depleted of
ices due to bombardment from Solar UV photons and cosmic rays
\citep{2004Icar..170..214M,2003EM&P...92..261C}.  These ionizing
particles are expected to destroy the molecular bonds of simple ices,
resulting in their depletion, and eventually yielding a dark, carbon
covered surface.  Indeed, many faint KBOs have been studied
spectroscopically, and in this work photometrically, the majority of
which are neutral in reflectance.  However, the presence of both
methane ice and water ice, although the minority for all KBOs, is
pervasive throughout the dynamical classes in the Kuiper Belt.  In our
survey, we find a Scattered KBO with ice (Eris), Resonant bodies
(Ixion and Orcus) and Classical (Makemake) bodies with ices as shown
in Table~\ref{familymembers}.  Among previously published
spectroscopic works the same pervasive nature of water ice is apparent
from Table~\ref{resultstable} and references therein, with Scattered
(Sedna and (26181)), Resonant ((47171), (55638), (84922) and (208996))
and Classical ((19308) and Quaoar) bodies.  Thus, our survey and
previous spectroscopic works independently suggest that ices are found
in among all orbital classes throughout the Kuiper Belt and are thus
not uniformly destroyed by solar UV photons and cosmic rays.  Nor does
it appear that ice surfaces are a extreme minority exception to the
radiation processing scenario, as roughly 10\% of the non-Haumea
bodies in our survey have ice of some kind on their surfaces.

\subsection{Sensitivity}
\label{sensitivity}

Our sensitivity to ices can be estimated by comparing the results of
our photometry with spectroscopic observations from literature, as
seen in Table~\ref{resultstable}.  For all of the Haumea family
members, which exhibit extremely large water ice absorption from 55\%
-- 80\% depth in absorption, our photometric work and spectroscopic
works agree.  For non-Haumea bodies, some of which show very low
levels of absorption ($\sim 15\%$ absorption depth), the results are
somewhat discrepant, which marks the minimum absorption levels to
which we are sensitive.  In addition, bodies with significant
quantities of ultra-red matter may have ices that are not easily
detectable with our $J$-band continuum estimation.

\subsubsection{Ultra-Red Matter and $J$-band Continuum Estimation}
\label{ultrared}

A key quantity for estimating the amount of absorption in a spectrum
is an accurate assessment of the continuum, i.e the reflectance found
outside of any absorptions.  For our survey, we have chosen the
$J$-band because it is generally unaffected by ices, as described in
\S\ref{filterdesign}.  Unfortunately, the presence of the so-called
``ultra-red matter'' \citep{2002AJ....123.1039J} can still impact this
estimate.  The ultra-red matter, which primarily colors the visible,
does extend partway into the $J$-band to about 1.3 microns.  The best
example of this spectroscopically is found for Quaoar
\citep{2004Natur.432..731J}, which is one of the reddest KBOs in the
visible (and thus has a large amount of ultra-red matter), but has a
relatively colorless near-IR continuum (attested by the fact that the
near-ir shape can be fit with a simple water ice model with no broad
coloring agent).  A careful examination of Quaoar's spectrum indicates
that the ultra-red matter likely extends into the $J$-band, although
the exact point where its absorption ends is not clear, since the
spectral shape of the ultra-red matter is unknown.  However, if we
assume that Quaoar's J-band spectrum differs from that of pure water
ice only due to the contribution of ultra-red matter, then the maximum
contamination that the ultra-red matter can impart to a $J$-band
continuum measurement is 0.15 magnitudes, which is significantly
smaller than the $\sim 1$ magnitude of color difference between
neutral and icy bodies in our experiement.

Thus, for Quaoar, and any other bodies with extremely red visible
slopes, the $J$-band continuum measurement could be depressed by 0.15
magnitudes using our method.  Thus, we use the $Y-J$ colors in our
survey to identify extremely red KBOs, which could harbor more ices on
their surfaces than our photometry indicates.  We have flagged these
bodies in Table~\ref{resultstable} by examining the $Y-J$ colors we
have collected.  We indicate red bodies as being those bodies with
$Y-J > 0.1$ to better than $3\sigma$ confidence.  Of the 29 bodies
with $Y-J$ known to better than 0.15 magnitudes, only 3 bodies have
such red colors: Quaoar, 2005 $\rm UQ_{513}$ and 2004 $\rm PF_{115}$.
Thus, the point of rejecting objects with extreme $Y-J$ values is that
red $Y-J$ color is a sign that the $J$-band flux measurement may be
contaminated by ultra-red matter, and therefore ice detection may be
depressed by not more than 0.15 magnitudes.

\subsubsection{Absorption Depth and Surface Ice Fraction}

Rejecting the reddest objects in the sample yields a consistent
picture between spectroscopy and photometry --- our survey can
identify objects with $\gtrsim 15\%$ water ice absorption depth.
Ixion and Orcus were the two non-Haumea family members identified as
having water ice by our photometric survey with band depths also
measured by spectroscopy.  They have $\sim 10\%$ and $\sim 20\%$ water
ice absorption depths respectively as found by
\cite{2005ApJ...627.1057T}, \cite{2005AA...437.1115D},
\cite{2008AA...479L..13B} and \cite{2010Icar..208..945M}.  Seven
bodies were suspected to have water ice in spectroscopic works, but
didn't indicate water in our work, as shown in
Table~\ref{resultstable}.  The mean water ice band depth for these
bodies is $12.5\% \pm 2.5\%$.  Thus, we estimate that for our mean
sample color accuracy of 0.1 magnitudes, we are sensitive to water ice
band depths $\gtrsim 15\%$.  Since the bodies with large methane
absorption show a similar $\sim 1$ magnitude separation from solar
bodies in our system, we expect similar sensitivities for the methane
band.

We also estimated our sensitivity by producing a very simple model of
a range of bodies from neutral, to pure water ice, to pure methane
ice, as shown in Figure~\ref{results}.  We use a simple bidirectional
reflectance model similar to that described in
\cite{2005ApJ...627.1057T,2007ApJ...655.1172T}, based on
\cite{1993tres.book.....H}.  Using absorption coefficients for water
ice \cite{1998JGR...10325809G} and methane ice
\cite{2002Icar..155..486G}, we computed the full-disk albedo expected
from a body covered in 1 mm diameter grains at 40 K.  We computed a
suite of models with these parameters, as presented as the grey
triangular matrix in Figure~\ref{results}.  The vertical scale bar
labeled ``Ice Fraction'' represents total ice percentage (methane ice
and water ice combined), with values less than 100\% indicating the
presence of a fictitious neutral absorber with albedo equal to the
mean albedo of KBOs, 0.1 \citep{2008ssbn.book..161S}.  The diagonal
scale bar labeled ``Methane Fraction'' indicates the percentage of
methane ice, with values less than 100\% including a linear
combination of water ice.  Thus, the top corner of the triangular grid
represents a neutral body, the lower rightmost corner represents a
body of 100\% methane ice, and the leftmost corner represents a body
of 100\% water ice.  Although many of the bodies in the graph appear
to fall outside of the grid, this is expected for our crude three
component model, as the real variety of surfaces on KBOs is certainly
much more complex.  The model does have utility, however, in that it
provides a rough constraint on the minimum surface fraction of ice to
which we are sensitive, roughly $\sim 10\%$ surface fraction in our
model, which is consistent with the $\sim 15\%$ water band depth for
both Orcus and Ixion.

\section{Discussion}
\label{discussion}

\subsection{Size versus Surface Type}
\label{size}

We see two clear trends in our data set: (1) the
surfaces of the smallest Haumea family members have a much stronger
incidence of water ice than Haumea itself and (2) only the largest
KBOs have methane ice.

\subsubsection{Methane on Large Bodies}

\begin{figure}[htbp]
\plotone{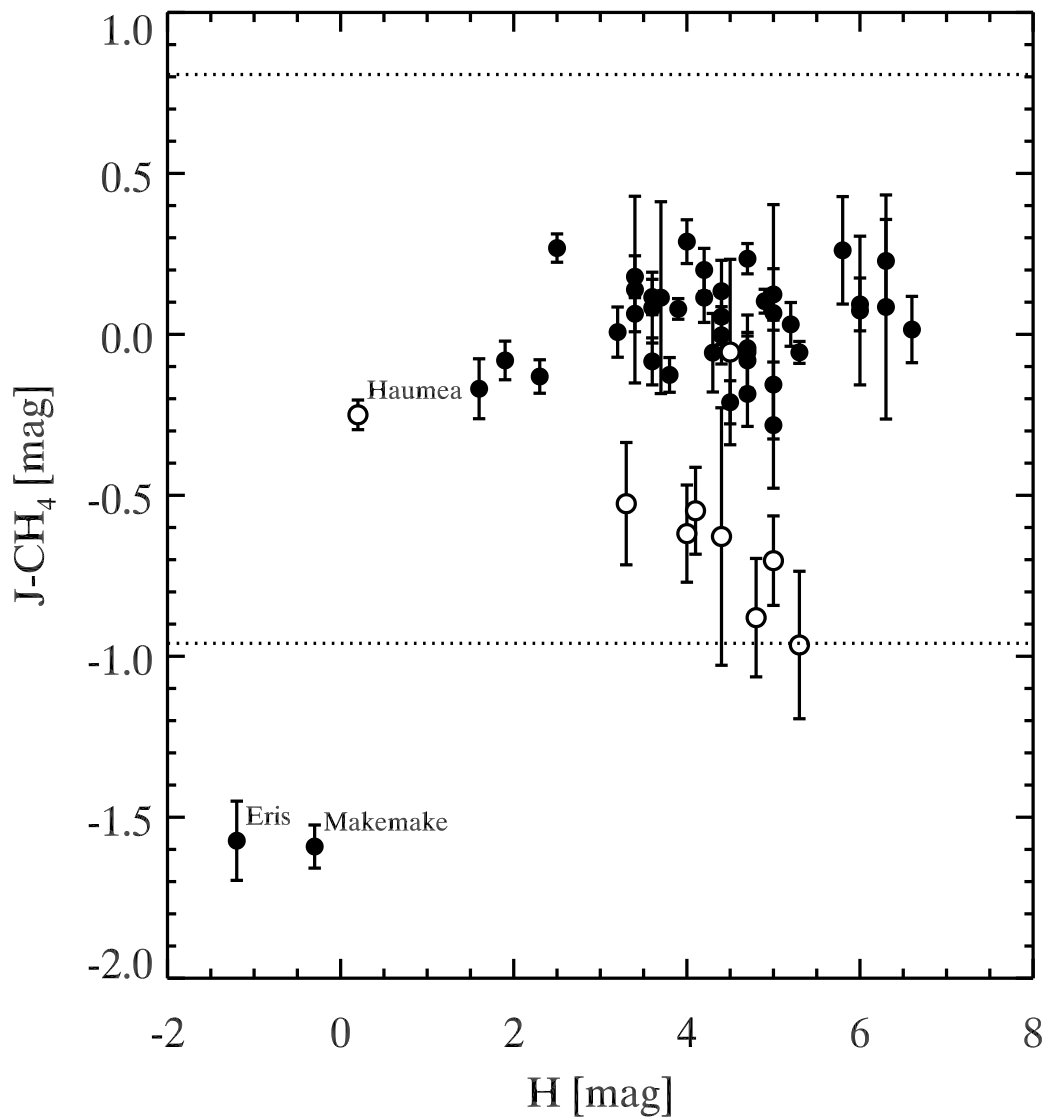}
\vspace*{0.5in} \figcaption{Methane ice as a function of size using
  the proxies $J-CH_{4}$ for methane ice fraction and $H$ (absolute
  magnitude) as a proxy for size.  Only the largest bodies in our
  sample, Makemake and Eris show evidence for methane ice.  The dotted
  lines represent the $\pm 3 \sigma$ limits to the non-methane bodies
  based upon their $J-CH_{4}$ values.  Hollow circles are Haumea and
  its family members.
\label{methanesize}}
\end{figure}

To date, there are only three bodies known in the Kuiper Belt with
strong methane detections, Makemake and Eris, as shown by our survey
in Figure~\ref{methanesize} and Pluto.  We use $H$ as a proxy for
size, which implicitly assumes a fixed albedo among the bodies,
because the vast majority of bodies in the plot have unknown albedos.
The possibility that only large bodies may harbor methane ice was
suggested by \cite{2007ApJ...659L..61S} and was used as the basis for
a simple model explaining surface phenomena based on vapor pressure.
Simply put, only very large bodies have surface gravity strong enough
to overcome methane vapor pressure even at large heliocentric
distances.  We tested this theory measuring a large number of bodies
with a variety of sizes with a single methodology.  We find no strong
evidence for methane on any bodies in our sample besides Makemake and
Eris.  Although both Quaoar and Sedna have been reported to have
alkanes on their surfaces, neither has a large enough amount of the
material to be detected in our survey
\citep{2004Natur.432..731J,2005AA...439L...1B,2007ApJ...670L..49S,2007DPS....39.4906T,2007A&A...466..395E}.
For our sample of 51 objects, the 2 with the smallest absolute
magnitudes (a proxy for largest size in the fixed albedo case) have
methane ice.  The probability that this could occur in size-neutral
process which coated 2 of 51 bodies with methane ice is given by the
binomial theorem as 0.12\%, thus this possibility can be rejected with
about $3.5 \sigma$ confidence.  To date, our observations are
consistent with the \cite{2007ApJ...659L..61S} theory.

\subsubsection{Water in the Haumea Family}

\begin{figure}[htbp]
\plotone{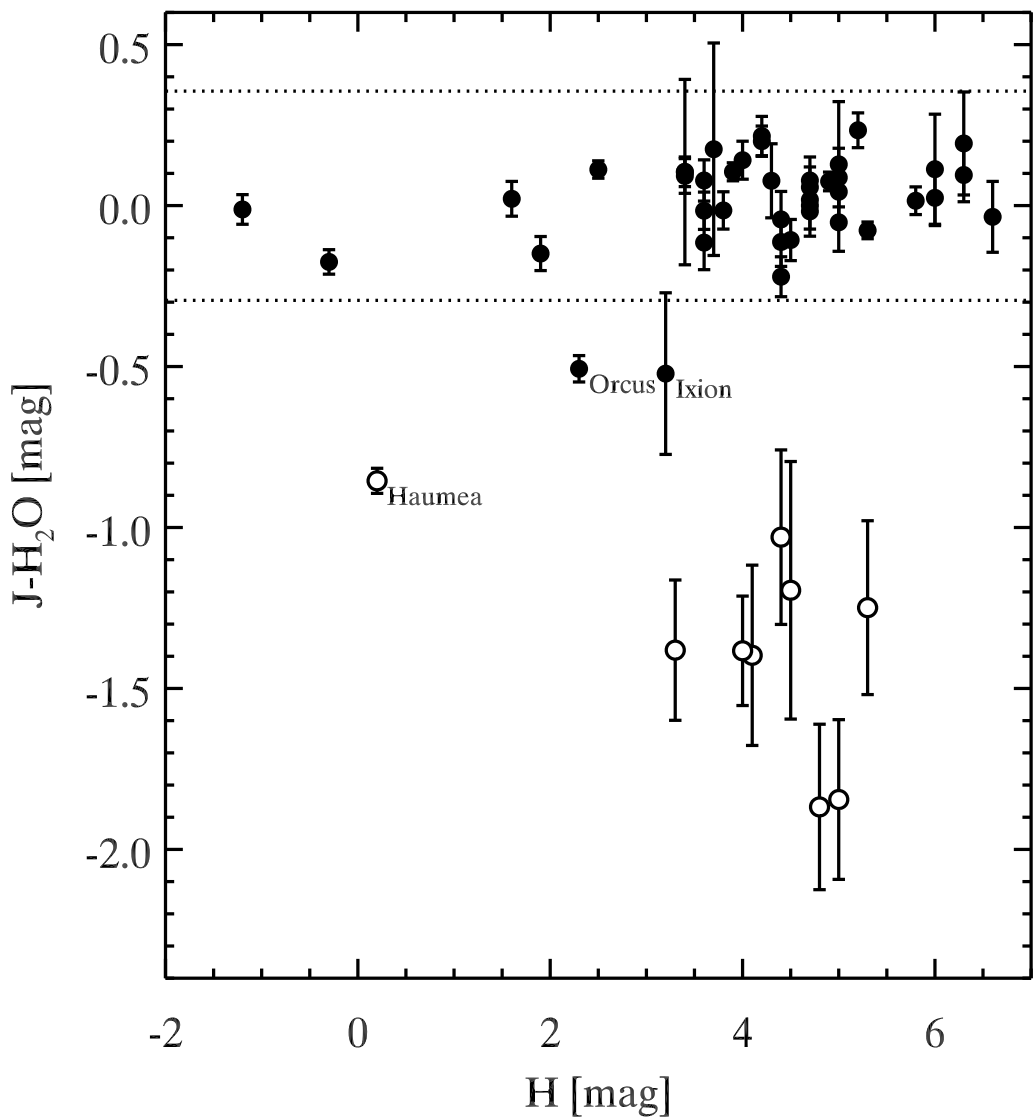}
\vspace*{0.5in} \figcaption{Water ice as a function of size using the
  proxies $J-H_{\rm 2}O$ for water ice fraction and $H$ (absolute
  magnitude) as a proxy for size.  The small Haumea family members
  show much larger water ice absorption than Haumea itself.  This
  could be explained by (1) a larger surface fraction covered in water
  ice or (2) larger grain sizes.  The dotted lines represent the $\pm
  3 \sigma$ limits to the non-family members based upon their
  $J-H_{\rm 2}O$ values.  Orcus and Ixion fall outside of these $3
  \sigma$ limits and we thus consider them to have significant amounts
  of ice on their surfaces.  Hollow circles are Haumea and its family
  members.
\label{watersize}}
\end{figure}

Examining water ice on the Haumea family members we find that water
ice presence is inversely correlated with size, or the smallest bodies
are the ones with the most water ice on their surface, as shown in
Figure~\ref{watersize}.  In this figure, we use absolute magnitude,
$H$, as a proxy for size since the albedos of Haumea's progeny are
unknown.  Using $H$ as a proxy for size implicitly assumes a fixed
albedo among the bodies, which may not be the case.  However, due to
the large magnitude difference ($\sim$ 3 magnitudes, or a factor 16 in
flux) between Haumea and its progeny, it is extremely unlikey that
even the brightest of the fainter bodies are larger than Haumea.  We
do not believe this to be the result of any observational bias, as our
standard star observations show no such correlation with apparent
magnitude.  We assess the statistical validity of this result by
showing that the Haumea $J-H_{2}O = -0.85 \pm 0.04$ with $H=0.2$ is
significantly different from the rest of the Haumea family, all of
which have $3.3 < H < 5.3$ and a population mean of $J-H_{2}O = -1.5
\pm 0.1$ (excluding $\rm 2009\ YE_{7}$).  Thus, Haumea has less water
absorption than the rest of the Haumea family with a confidence level
of $\sim 6\sigma$.

It is not clear the underlying physical cause of such a difference,
however, there are two obvious possibilities: (1) the smaller bodies
have a larger fraction of water ice on their surfaces or (2) the
smaller bodies have a larger grain size, leading to stronger
absorption features.  Both of these possibilities are physically
plausible.  If we assume that the Haumea family was created in a
collisional event involving a much larger KBO, it is very likely that
this body had some form of differentiation
\citep{2008ssbn.book..213M}.  Thus, the largest remaining member,
Haumea, likely contains a larger fraction of rock in its core than the
smaller family members which were created from the ejecta material,
which was presumably closer to the surface, less dense, and with less
structural strength than the Haumea material.  Thus it is easily
plausible that scenario (1) is valid because the family members
contain more water ice on a bulk level.  Scenario (2) is also
plausible since any ice transport is likely very strongly correlated
with surface gravity.  If significant surface evolution has taken
place on the Haumea family, which has been suggested as a possible
explanation for the crystalline water ice presence on their surfaces
\citep{2008Icar..193..397N}, then it is very likely that this process
is affected by surface gravity.  Surface gravity profoundly affects
ice transport and geologic processes such as cryovolcanism, thus
nearly any evolutional process whether is be external
(e.g. micro-meteorite bombardment leading to ice transport
\citep{2010Icar..208..492P}) or internal (e.g. cryovolcanism
\citep{2004Natur.432..731J}) in origin likely is affected by body
size.  It should be noted that although cryovolcanism may be expected
for larger bodies, in-situ observations of the $\sim 2700$ km diameter
Triton by Voyager show only a small amount of surface is likely
affected by cryovolcanism \citep{1989Sci...246.1369A}.  Recent
theoretical modeling shows that cryovolcanism in large ($\gtrsim 1200$
km diameter) KBOs may be enhanced over that of Triton
\citep{2009Icar..202..694D}, making the role of cryovolcanism in
disk-integrated surface properties still an open question.

\subsection{The Role of Photometry in Future Work}
\label{advantages}

Our custom photometric survey is not meant to replace spectroscopy in
any way.  The limited spectral information provided in a photometric
survey covering only a few filters doesn't provide anywhere near the
amount of information needed to disentangle the astrophysical
processes taking place on minor bodies in the outer solar system.
Indeed, in-situ observations of Saturnian satellites such as Iapetus,
Phoebe and Hyperion show how remarkably diverse small body surfaces
are in the outer solar system with extremely high signal-to-noise
measurements.  Thus, our survey is not meant to be an ending point of
Kuiper Belt surface study but a starting point that will identify the
bodies of greatest interest for later follow-up work.

\begin{figure}[htbp]
\epsscale{0.75}
\plotone{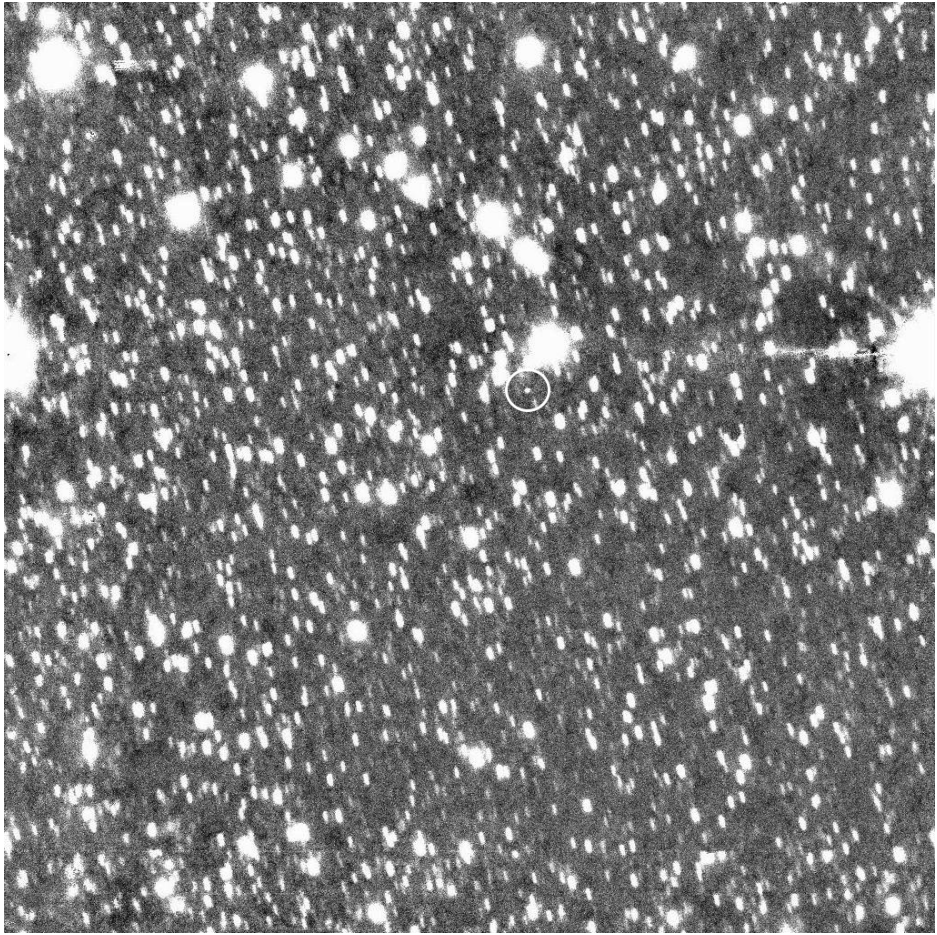}
\vspace*{0.5in} \figcaption{An example of a rich field from Magellan
  appears above.  The KBO $\rm 2002\ MS_{4}$ is shown in a deep
  composite image of all frames taken over a $\sim 2$ hour period,
  digitally tracked at the non-sidereal rate of the object.  A deep
  image was constructed of every science field in our sample to ensure
  no stellar contamination.  Such a verification procedure is not
  possible for traditional spectroscopic measurements and background
  stars can easily unknowingly contaminate observations.
\label{crowdedfield}}
\end{figure}

Photometry does offer several distinct advantages over spectroscopy:

\begin{enumerate}

\item Numerous instrumental losses are associated with spectroscopy.
  Although each instrumental setup is different, observations from
  Gemini North telescope are fairly typical for near-infrared
  instruments.  Losses associated with an hour of NIRI spectroscopy
  would typically include the following: 20\% loss from light near the
  periphery of the object that is occulted by the slit, 10\%
  acquisition overhead required to put the target in the slit, a 50\%
  loss due to grism throughput and image quality degradation through
  the grism.  In addition, spectral information is spread across a few
  thousand pixels of the detector, whereas photometric information is
  tightly associated in twenty pixels, greatly reducing instrumental
  noise.

\item Sky line removal can be difficult for spectral integrations
  which often require $\sim 5$ minute exposures for faint object
  signal to be apparent.  Since OH lines emission can typically change
  on $\sim 5$ minute timescales, integrations longer than this can
  make sky line removal challenging.  Since photometric integrations
  are typically of order $\sim 1$ minute, pairwise subtraction of
  images efficiently removes most of the changing sky background.

\item In crowded fields, potential contamination of bodies can be
  easily identified, as depicted in Figure~\ref{crowdedfield}.
  Spectroscopic observations typically only spend $\sim 10$ minutes
  imaging the science target during acquisition, with spectroscopic
  observations taking the vast majority of the telescope time.  Thus,
  it is impossible to construct a deep-field image of the science
  target environment when performing spectroscopy, which, especially
  for objects near the galactic plane, can contain many contaminant
  background stars.

\end{enumerate}

Because of these advantages, we find that we can save roughly a factor
$\sim 3$ in telescope time while still identifying the objects of key
interest, such as the Haumea family members.  This gives us a large
potential increase in sample size, and allows us to observe many more
objects with a consistent instrumental setup in consistent conditions
than an equivalent spectroscopic work.  In addition, a population-wide
study such as ours has the ability to identify important population
outliers for further spectroscopic study by the sheer number of
objects available in our sample.

In this work, our primary goal is to present the details of the custom
photometric filter system.  A future work will include a must larger
sample of bodies, including faint bodies that have to date been out of
reach of spectroscopic works.  Any bodies with unusual surface
features will be identified for further study in a subsequent work.
Since many of these bodies will be fainter than the $V \sim 22$
brightness level, they will require $10 \sim 30$ hours of 10 meter
telescope time for follow-up spectroscopy to determine more about
their surfaces.  Requesting this amount of telescope time for a single
object with no prior information about its surface is simply not
feasible.  Nor is it a wise use of telescope resources when most faint
bodies appear to be bland.  Thus, we believe that future results from
our survey will become the cornerstone of spectroscopic study of faint
bodies in the upcoming decade and as 30 meter class telescopes begin
to come on line.

\section{Summary}

The primary purpose of this work is to detail our custom photometric
system and demonstrate that it is useful in identifying dynamical
correlations on the subset of KBOs we have observed to date.  We find
the following in this work.

\begin{enumerate}

\item Our custom photometric work can identify KBO surfaces with a
  factor $\sim 3$ less telescope time than similar quality
  spectroscopic works.  We can discriminate neutral bodies from water
  and methane ice bodies with $\gtrsim 15\%$ absorption depths, which
  corresponds to approximately $\gtrsim 10\%$ surface fractions of
  ices for simple models of KBO surfaces.

\item We identify 9 Haumea family members including a new object 2009
  $\rm YE_{7}$, which is consistent in visible color, water ice depth
  and proper orbital elements with the rest of the Haumea family.  We
  reject several objects with dynamics similar to Haumea because they
  do not have water ice on their surfaces.

\item We find that the smallest Haumea family members systematically
  have deeper water ice absorption than Haumea to $6 \sigma$
  confidence.  This observation is consistent with a collisional
  origin for the system.

\item We identify two non-Haumea bodies with water ice in 3:2
  mean-motion resonance with Neptune (Orcus and Ixion).  Combining
  this with previous spectroscopic studies demonstrates that moderate
  water ice is pervasive throughout all Kuiper Belt dynamical classes.

\item Outside the Haumea family, we find no evidence for any
  correlation between body size and water ice fraction.

\item We find that only the largest KBOs harbor methane ice, which is
  consistent with arguments of volatile loss timescales based on
  surface gravity, such as those presented by
  \cite{2007ApJ...659L..61S}.

\end{enumerate}

\acknowledgements

We thank the anonymous reviewer for providing constructive comments
which improved this work.  Gemini observations were conducted under
program IDs GN-2008A-Q-4, GN-2008B-Q-40, GN-2009A-Q-6, and
GN-2009B-Q-30.  This work was supported in part by NASA Planetary
Astronomy Grant NNX07AK96G.

\bibliography{myrefs}
\bibliographystyle{apj}

\clearpage
\begin{deluxetable}{lllll}
\tablewidth{3.5in}
\tablecaption{Filter Wavelength Definitions}
\tablehead{
\colhead{Filter} & \colhead{$\lambda_{\rm min}$} &
\colhead{$\lambda_{\rm max}$} & \colhead{$\lambda_{\rm central}$} & \colhead{$\Delta \lambda$} \\
\colhead{} & \colhead{[\micron]} & \colhead{[\micron]} & \colhead{[\micron]} & \colhead{[\micron]} }
\startdata
Water   & 1.9490 & 2.1195 & 2.0343 & 0.1705 \\
Methane & 2.1990 & 2.3505 & 2.2748 & 0.1515 \\
\enddata
\label{filterdetails}
\tablecomments{Wavelengths are defined to be the wavelength where the
  filter transmissions reached 50\% of the maximum transmission.
  Minimum and maximum wavelengths are given by $\lambda_{\rm min}$ and
  $\lambda_{\rm max}$, respectively while central wavelength and
  filter widths are given by $\lambda_{\rm central}$ and $\Delta \lambda$,
  respectively. \\ }
\end{deluxetable}

\clearpage
\begin{deluxetable}{rllrrrrrrrl}
\rotate
\tabletypesize{\scriptsize}
\tablewidth{8.5in}
\tablecolumns{11}
\tablecaption{Observational Circumstances}
\tablehead{
\colhead{Object} & \colhead{Object} & \colhead{Provisional} & \colhead{Site} & \colhead{UT Date} & \colhead{UT Time} & \colhead{Airmass} & \colhead{J}   & \colhead{$\rm H_{2}O$} & \colhead{$\rm CH_{4}$} & \colhead{Calibration} \\
\colhead{Number} & \colhead{Name}   & \colhead{Designation} & \colhead{}     & \colhead{}        & \colhead{Range}   & \colhead{Range}   & \colhead{[s]} & \colhead{[s]}          & \colhead{[s]}         & \colhead{Sources} }
\startdata
         &                & $ \rm 2000\ CN_{105} $ & Gemini   & 2009-02-01 & 12:25 -- 13:37 & 1.014 -- 1.065 &  450 & 1440 & 1320 & HIP24336, HIP55398 \\
         &                & $ \rm 2001\ KD_{77} $  & Gemini   & 2009-06-15 & 12:16 -- 13:06 & 1.560 -- 1.887 &  150 &  360 &  240 & HIP90869 \\
         &                & $ \rm 2001\ QC_{298} $ & Gemini   & 2009-08-04 & 12:29 -- 13:27 & 1.049 -- 1.078 &  180 &  780 &  420 & HIP111063, HIP117367, HIP9829 \\
         &                & $ \rm 2002\ GH_{32} $  & Gemini   & 2009-01-31 & 15:30 -- 16:08 & 1.223 -- 1.259 &  270 &  660 &  720 & HIP23259, HIP49580, HIP49942 \\
         &                & $ \rm 2002\ KW_{14} $  & Gemini   & 2009-06-30 & 09:34 -- 10:34 & 1.515 -- 1.887 &  360 & 1080 & 1080 & HIP78107, HIP80609 \\
         &                & $ \rm 2002\ MS_{4} $   & Magellan & 2009-04-05 & 09:09 -- 10:34 & 1.066 -- 1.113 &  525 &  600 & 1170 & HD110747, HD118928, HD138159 \\
         &                &                       &          &            &                &                &      &      &      & HD142801, HD170717, HD95868 \\
         &                & $ \rm 2002\ XV_{93} $  & Gemini   & 2009-01-31 & 10:05 -- 10:43 & 1.455 -- 1.715 &  270 &  660 &  720 & HIP23259, HIP49580, HIP49942 \\
         &                & $ \rm 2003\ FE_{128} $ & Gemini   & 2009-04-15 & 12:33 -- 13:13 & 1.241 -- 1.336 &  270 &  720 &  600 & HIP73606 \\
         &                & $ \rm 2003\ UZ_{117} $ & Gemini   & 2009-09-20 & 13:13 -- 14:11 & 1.032 -- 1.049 &  360 &  540 &  960 & HIP19767, HIP7373 \\
         &                & $ \rm 2003\ UZ_{117} $ & Gemini   & 2009-12-30 & 06:11 -- 07:07 & 1.035 -- 1.071 &  360 &  960 & 1020 & HIP117367, HIP18768, HIP19767 \\
         &                & $ \rm 2003\ UZ_{413} $ & Gemini   & 2009-01-09 & 07:18 -- 08:01 & 1.105 -- 1.205 &  270 &  660 &  960 & HIP19767 \\
         &                & $ \rm 2004\ NT_{33} $  & Gemini   & 2009-06-24 & 12:55 -- 13:50 & 1.004 -- 1.025 &  360 & 1080 & 1080 & HIP106356 \\
         &                & $ \rm 2004\ NT_{33} $  & Gemini   & 2009-08-06 & 10:39 -- 11:36 & 1.012 -- 1.074 &  330 &  960 & 1080 & HIP106356 \\
         &                & $ \rm 2004\ PT_{107} $ & Gemini   & 2009-09-10 & 06:00 -- 06:37 & 1.411 -- 1.603 &  270 &  660 &  540 & HIP111063, HIP117367 \\
         &                & $ \rm 2005\ CB_{79} $  & Gemini   & 2009-01-31 & 11:47 -- 12:26 & 1.019 -- 1.054 &  270 &  600 &  660 & HIP23259, HIP49580, HIP49942 \\
         &                & $ \rm 2005\ GE_{187} $ & Gemini   & 2009-01-31 & 13:58 -- 14:36 & 1.427 -- 1.655 &  270 &  720 &  420 & HIP23259, HIP49580, HIP49942 \\
         &                & $ \rm 2005\ QU_{182} $ & Gemini   & 2009-09-20 & 08:40 -- 09:57 & 1.218 -- 1.499 &  390 & 1440 & 1380 & HIP19767, HIP7373 \\
         &                & $ \rm 2007\ JH_{43} $  & Magellan & 2009-04-04 & 07:33 -- 09:19 & 1.022 -- 1.078 &  600 & 1320 & 1935 & HD123385, HD138159, HD154805 \\
         &                &                       &          &            &                &                &      &      &      & HD77533, HD85538 \\
         &                & $ \rm 2008\ LP_{17} $  & Gemini   & 2009-06-29 & 07:39 -- 08:51 & 1.183 -- 1.361 &  450 & 1440 & 1380 & HIP113050, HIP75923 \\
         &                & $ \rm 2009\ YE_{7} $   & Gemini   & 2009-12-30 & 07:20 -- 07:59 & 1.146 -- 1.154 &  270 &  480 &  660 & HIP117367, HIP18768, HIP19767 \\
         &                & $ \rm 2009\ YE_{7} $   & Gemini   & 2010-02-21 & 05:42 -- 06:35 & 1.231 -- 1.405 &  330 &  720 &  420 & HIP18768 \\
 (19308) &                & $ \rm 1996\ TO_{66} $  & Gemini   & 2009-09-18 & 09:56 -- 10:45 & 1.011 -- 1.048 &  270 &  660 &  660 & HIP117367, HIP11747, HIP9829 \\
 (19308) &                & $ \rm 1996\ TO_{66} $  & Gemini   & 2009-12-30 & 05:23 -- 06:03 & 1.058 -- 1.132 &  270 &  600 &  540 & HIP117367, HIP18768, HIP19767 \\
 (24835) &                & $ \rm 1995\ SM_{55} $  & Gemini   & 2009-09-18 & 13:13 -- 14:07 & 1.007 -- 1.044 &  270 & 1260 &  660 & HIP117367, HIP11747, HIP9829 \\
 (26181) &                & $ \rm 1996\ GQ_{21} $  & Gemini   & 2009-06-30 & 07:49 -- 09:01 & 1.193 -- 1.395 &  450 & 1440 & 1260 & HIP78107, HIP80609 \\
 (26375) &                & $ \rm 1999\ DE_{9} $   & Magellan & 2009-04-05 & 02:09 -- 04:07 & 1.140 -- 1.223 &  450 & 1980 & 1665 & HD110747, HD118928, HD138159 \\
         &                &                       &          &            &                &                &      &      &      & HD142801, HD170717, HD95868 \\
 (28978) &          Ixion & $ \rm 2001\ KX_{76} $  & Gemini   & 2009-05-01 & 14:43 -- 15:03 & 1.664 -- 1.790 &  180 &  300 &  300 & HIP83875, HIP84181 \\
 (38628) &           Huya & $ \rm 2000\ EB_{173} $ & Magellan & 2009-04-05 & 05:06 -- 05:59 & 1.152 -- 1.263 &  300 &  600 &  990 & HD110747, HD118928, HD138159 \\
         &                &                       &          &            &                &                &      &      &      & HD142801, HD170717, HD95868 \\
 (40314) &                & $ \rm 1999\ KR_{16} $  & Gemini   & 2009-01-31 & 14:47 -- 15:26 & 1.185 -- 1.263 &  270 &  720 &  660 & HIP23259, HIP49580, HIP49942 \\
 (47171) &                & $ \rm 1999\ TC_{36} $  & Magellan & 2008-10-18 & 04:03 -- 05:03 & 1.129 -- 1.148 &  450 &  810 &  840 & HD9729, HIP11747 \\
 (47932) &                & $ \rm 2000\ GN_{171} $ & Gemini   & 2009-05-16 & 10:44 -- 11:24 & 1.384 -- 1.550 &  270 &  720 &  660 & HIP73606 \\
 (50000) &         Quaoar & $ \rm 2002\ LM_{60} $  & Magellan & 2009-04-05 & 07:23 -- 07:54 & 1.111 -- 1.181 &  300 &  300 &  300 & HD110747, HD118928, HD138159 \\
         &                &                       &          &            &                &                &      &      &      & HD142801, HD170717, HD95868 \\
 (55565) &                & $ \rm 2002\ AW_{197} $ & Gemini   & 2009-01-31 & 12:40 -- 13:18 & 1.146 -- 1.261 &  270 &  660 &  720 & HIP23259, HIP49580, HIP49942 \\
 (55565) &                & $ \rm 2002\ AW_{197} $ & Magellan & 2009-04-04 & 02:06 -- 02:57 & 1.229 -- 1.312 &  450 &  660 &  660 & HD123385, HD138159, HD154805 \\
         &                &                       &          &            &                &                &      &      &      & HD77533, HD85538 \\
 (55636) &                & $ \rm 2002\ TX_{300} $ & Gemini   & 2008-11-06 & 06:49 -- 07:21 & 1.015 -- 1.037 &  180 &  720 &  300 & HIP394 \\
 (55636) &                & $ \rm 2002\ TX_{300} $ & Gemini   & 2009-01-10 & 04:35 -- 04:53 & 1.041 -- 1.065 &  360 &  300 &  300 & HIP394 \\
 (55637) &                & $ \rm 2002\ UX_{25} $  & Magellan & 2008-10-18 & 05:24 -- 06:24 & 1.286 -- 1.366 &  450 &  540 & 1080 & HD9729, HIP11747 \\
 (55638) &                & $ \rm 2002\ VE_{95} $  & Gemini   & 2009-02-01 & 07:36 -- 08:31 & 1.127 -- 1.304 &  360 & 1080 & 1080 & HIP24336, HIP55398 \\
 (65489) &           Ceto & $ \rm 2003\ FX_{128} $ & Gemini   & 2009-05-13 & 08:00 -- 08:52 & 1.061 -- 1.129 &  240 &  780 &  660 & HIP72855, HIP82233 \\
 (84522) &                & $ \rm 2002\ TC_{302} $ & Gemini   & 2009-08-04 & 13:55 -- 14:36 & 1.028 -- 1.086 &  270 &  720 &  720 & HIP111063, HIP117367, HIP9829 \\
 (84922) &                & $ \rm 2003\ VS_{2} $   & Gemini   & 2008-12-02 & 10:49 -- 11:27 & 1.081 -- 1.147 &  270 &  720 &  720 & HIP21333 \\
 (86177) &                & $ \rm 1999\ RY_{215} $ & Gemini   & 2009-09-19 & 11:08 -- 12:23 & 1.225 -- 1.595 &  420 & 1260 & 1260 & HIP117367 \\
 (90377) &          Sedna & $ \rm 2003\ VB_{12} $  & Gemini   & 2008-12-03 & 09:54 -- 11:06 & 1.060 -- 1.209 &  450 & 1440 & 1440 & HIP18768 \\
 (90482) &          Orcus & $ \rm 2004\ DW_{} $    & Magellan & 2009-04-04 & 00:32 -- 01:40 & 1.091 -- 1.159 &  300 & 1080 & 1080 & HD123385, HD138159, HD154805 \\
         &                &                       &          &            &                &                &      &      &      & HD77533, HD85538 \\
 (90568) &                & $ \rm 2004\ GV_{9} $   & Magellan & 2009-04-04 & 06:10 -- 07:15 & 1.001 -- 1.032 &  900 &  300 &  480 & HD123385, HD138159, HD154805 \\
         &                &                       &          &            &                &                &      &      &      & HD77533, HD85538 \\
(119951) &                & $ \rm 2002\ KX_{14} $  & Gemini   & 2009-05-01 & 13:22 -- 14:09 & 1.441 -- 1.629 &  270 &  720 &  720 & HIP83875, HIP84181 \\
(120132) &                & $ \rm 2003\ FY_{128} $ & Magellan & 2009-04-04 & 03:05 -- 04:11 & 1.067 -- 1.163 &  450 &  660 & 1260 & HD123385, HD138159, HD154805 \\
         &                &                       &          &            &                &                &      &      &      & HD77533, HD85538 \\
(120178) &                & $ \rm 2003\ OP_{32} $  & Magellan & 2008-10-18 & 01:23 -- 03:42 & 1.205 -- 1.727 &  930 & 1200 & 2220 & HD9729, HIP11747 \\
(120347) &                & $ \rm 2004\ SB_{60} $  & Gemini   & 2008-11-28 & 04:53 -- 06:08 & 1.006 -- 1.085 &  450 & 1440 & 1440 & HIP113688 \\
(120348) &                & $ \rm 2004\ TY_{364} $ & Gemini   & 2009-09-18 & 12:00 -- 12:40 & 1.167 -- 1.196 &  270 &  720 &  720 & HIP117367, HIP11747, HIP9829 \\
(136108) &         Haumea & $ \rm 2003\ EL_{61} $  & Magellan & 2009-04-05 & 04:24 -- 04:50 & 1.574 -- 1.652 &  150 &  300 &  450 & HD110747, HD118928, HD138159 \\
         &                &                       &          &            &                &                &      &      &      & HD142801, HD170717, HD95868 \\
(136108) &         Haumea & $ \rm 2003\ EL_{61} $  & Magellan & 2009-04-04 & 05:22 -- 05:52 & 1.505 -- 1.525 &  300 &  300 &  300 & HD123385, HD138159, HD154805 \\
         &                &                       &          &            &                &                &      &      &      & HD77533, HD85538 \\
(136199) &           Eris & $ \rm 2003\ UB_{313} $ & Magellan & 2008-10-18 & 06:30 -- 07:16 & 1.215 -- 1.371 &  300 &  300 & 1020 & HD9729, HIP11747 \\
(136472) &       Makemake & $ \rm 2005\ FY_{9} $   & Magellan & 2009-04-04 & 04:34 -- 05:05 & 1.881 -- 1.891 &  150 &  300 &  510 & HD123385, HD138159, HD154805 \\
         &                &                       &          &            &                &                &      &      &      & HD77533, HD85538 \\
(145451) &                & $ \rm 2005\ RM_{43} $  & Magellan & 2008-10-18 & 07:36 -- 09:09 & 1.213 -- 1.445 &  750 & 1260 & 1380 & HD9729, HIP11747 \\
(145451) &                & $ \rm 2005\ RM_{43} $  & Magellan & 2008-10-19 & 08:43 -- 09:32 & 1.359 -- 1.584 &  300 &  810 & 1080 & HIP18768 \\
(145452) &                & $ \rm 2005\ RN_{43} $  & Gemini   & 2008-11-01 & 05:22 -- 06:36 & 1.067 -- 1.116 &  450 & 1440 & 1440 & HIP110035, HIP394 \\
(145453) &                & $ \rm 2005\ RR_{43} $  & Magellan & 2008-10-19 & 04:15 -- 05:58 & 1.203 -- 1.498 &  600 & 1890 & 1260 & HIP18768 \\
(145480) &                & $ \rm 2005\ TB_{190} $ & Gemini   & 2009-08-07 & 09:34 -- 10:48 & 1.119 -- 1.338 &  420 & 1440 & 1380 & HIP117367, HIP80609 \\
(145480) &                & $ \rm 2005\ TB_{190} $ & Gemini   & 2009-09-10 & 07:12 -- 08:26 & 1.132 -- 1.371 &  450 & 1440 & 1440 & HIP111063, HIP117367 \\
(174567) &                & $ \rm 2003\ MW_{12} $  & Magellan & 2009-04-05 & 06:17 -- 07:16 & 1.235 -- 1.451 &  450 &  660 &  990 & HD110747, HD118928, HD138159 \\
         &                &                       &          &            &                &                &      &      &      & HD142801, HD170717, HD95868 \\
(175113) &                & $ \rm 2004\ PF_{115} $ & Gemini   & 2009-08-08 & 10:56 -- 12:12 & 1.378 -- 1.407 &  450 & 1440 & 1440 & HIP110512 \\
(202421) &                & $ \rm 2005\ UQ_{513} $ & Gemini   & 2008-11-01 & 07:04 -- 08:19 & 1.009 -- 1.023 &  450 & 1440 & 1440 & HIP110035, HIP394 \\
(208996) &                & $ \rm 2003\ AZ_{84} $  & Gemini   & 2009-01-31 & 11:01 -- 11:39 & 1.124 -- 1.236 &  270 &  720 &  720 & HIP23259, HIP49580, HIP49942 \\
(208996) &                & $ \rm 2003\ AZ_{84} $  & Magellan & 2009-04-05 & 00:23 -- 01:53 & 1.359 -- 1.630 &  600 & 1320 & 1440 & HD110747, HD118928, HD138159 \\
         &                &                       &          &            &                &                &      &      &      & HD142801, HD170717, HD95868 \\
(225088) &                & $ \rm 2007\ OR_{10} $  & Gemini   & 2009-06-29 & 12:38 -- 13:50 & 1.213 -- 1.321 &  450 & 1440 & 1380 & HIP113050, HIP75923 \\
(225088) &                & $ \rm 2007\ OR_{10} $  & Gemini   & 2009-09-16 & 08:10 -- 09:25 & 1.212 -- 1.239 &  450 & 1380 & 1440 & HIP113050 \\
(229762) &                & $ \rm 2007\ UK_{126} $ & Magellan & 2008-10-19 & 06:23 -- 08:23 & 1.113 -- 1.184 &  360 & 2250 & 1380 & HIP18768 \\
\enddata
\label{observations}
\end{deluxetable}

\clearpage
\begin{deluxetable}{rllrrrlr}
\rotate
\tabletypesize{\footnotesize}
\tablewidth{7.5in}
\tablecolumns{8}
\tablecaption{Photometry}
\tablehead{
\colhead{Object} & \colhead{Object} & \colhead{Provisional} & \colhead{$J$}   & \colhead{$J - H_{2}O$} & \colhead{$J - CH_{4}$} & \colhead{Photometric} & \colhead{Spectroscopic} \\
\colhead{Number} & \colhead{Name}   & \colhead{Designation} & \colhead{[mag]} & \colhead{[mag]}       & \colhead{[mag]}        & \colhead{Type}        & \colhead{Absorption\tablenotemark{a}}}
\startdata
         &                & $ \rm 2000\ CN_{105} $ & $ 20.48 \pm 0.06 $ & $  0.09 \pm 0.09 $ & $ -0.28 \pm 0.20 $ & Neutral & \\
         &                & $ \rm 2002\ KW_{14} $  & $ 19.69 \pm 0.06 $ & $ -0.05 \pm 0.09 $ & $ -0.16 \pm 0.17 $ & Neutral & \\
         &                & $ \rm 2002\ MS_{4} $   & $ 19.71 \pm 0.27 $ & $  0.18 \pm 0.33 $ & $  0.11 \pm 0.30 $ & Neutral & \\
         &                & $ \rm 2002\ XV_{93} $  & $ 19.80 \pm 0.17 $ & $  0.13 \pm 0.19 $ & $  0.07 \pm 0.34 $ & Neutral & \\
         &                & $ \rm 2003\ FE_{128} $ & $ 20.54 \pm 0.13 $ & $  0.19 \pm 0.16 $ & $  0.09 \pm 0.35 $ & Neutral & \\
         &                & $ \rm 2003\ UZ_{117} $ & $ 20.27 \pm 0.07 $ & $ -1.25 \pm 0.27 $ & $ -0.96 \pm 0.23 $ &   Water\tablenotemark{f} & $\sim 80\%$ Water\tablenotemark{2} \\
         &                & $ \rm 2003\ UZ_{413} $ & $ 19.59 \pm 0.07 $ & $  0.08 \pm 0.11 $ & $ -0.06 \pm 0.12 $ & Neutral & \\
         &                & $ \rm 2004\ NT_{33} $  & $ 19.17 \pm 0.04 $ & $ -0.11 \pm 0.08 $ & $ -0.00 \pm 0.09 $ & Neutral & \\
         &                & $ \rm 2004\ PT_{107} $ & $ 20.23 \pm 0.11 $ & $  0.11 \pm 0.17 $ & $  0.07 \pm 0.23 $ & Neutral & \\
         &                & $ \rm 2005\ CB_{79} $  & $ 19.61 \pm 0.04 $ & $ -1.84 \pm 0.25 $ & $ -0.70 \pm 0.14 $ &   Water\tablenotemark{f} & $\sim 70\%$ Water\tablenotemark{2} \\
         &                & $ \rm 2005\ QU_{182} $ & $ 19.66 \pm 0.28 $ & $  0.10 \pm 0.29 $ & $  0.14 \pm 0.29 $ & Neutral & \\
         &                & $ \rm 2007\ JH_{43} $  & $ 19.36 \pm 0.04 $ & $  0.08 \pm 0.07 $ & $ -0.18 \pm 0.10 $ & Neutral & \\
         &                & $ \rm 2008\ LP_{17} $  & $ 19.74 \pm 0.06 $ & $ -0.03 \pm 0.11 $ & $  0.01 \pm 0.10 $ & Neutral & \\
         &                & $ \rm 2009\ YE_{7} $   & $ 20.84 \pm 0.11 $ & $ -1.03 \pm 0.27 $ & $ -0.63 \pm 0.40 $ &   Water & \\
 (19308) &                & $ \rm 1996\ TO_{66} $  & $ 20.73 \pm 0.23 $ & $ -1.20 \pm 0.40 $ & $ -0.06 \pm 0.29 $ &   Water\tablenotemark{f} & Water\tablenotemark{12,14} \\
 (24835) &                & $ \rm 1995\ SM_{55} $  & $ 19.43 \pm 0.05 $ & $ -1.87 \pm 0.26 $ & $ -0.88 \pm 0.18 $ &   Water\tablenotemark{f} & 56\% Water\tablenotemark{1} \\
 (26181) &                & $ \rm 1996\ GQ_{21} $  & $ 19.12 \pm 0.04 $ & $  0.23 \pm 0.05 $ & $  0.03 \pm 0.07 $ & Neutral &  9\% Water\tablenotemark{1} \\
 (26375) &                & $ \rm 1999\ DE_{9} $   & $ 19.25 \pm 0.05 $ & $ -0.02 \pm 0.08 $ & $ -0.05 \pm 0.11 $ & Neutral & Neutral\tablenotemark{1,3} \\
 (28978) &          Ixion & $ \rm 2001\ KX_{76} $  & $ 18.35 \pm 0.07 $ & $ -0.52 \pm 0.25 $ & $  0.01 \pm 0.08 $ &   Water & $\sim 10\%$ Water\tablenotemark{3,13} \\
 (38628) &           Huya & $ \rm 2000\ EB_{173} $ & $ 18.29 \pm 0.03 $ & $  0.06 \pm 0.06 $ & $  0.24 \pm 0.05 $ & Neutral & Neutral\tablenotemark{1,11} \\
 (40314) &                & $ \rm 1999\ KR_{16} $  & $ 19.48 \pm 0.03 $ & $  0.01 \pm 0.04 $ & $  0.26 \pm 0.17 $ & Neutral & \\
 (47171) &                & $ \rm 1999\ TC_{36} $  & $ 18.00 \pm 0.02 $ & $  0.07 \pm 0.03 $ & $  0.10 \pm 0.04 $ & Neutral & 8\% Water\tablenotemark{1} \\
 (47932) &                & $ \rm 2000\ GN_{171} $ & $ 19.07 \pm 0.04 $ & $  0.02 \pm 0.09 $ & $  0.09 \pm 0.08 $ & Neutral & Neutral\tablenotemark{1} \\
 (50000) &         Quaoar & $ \rm 2002\ LM_{60} $  & $ 17.25 \pm 0.02 $ & $  0.11 \pm 0.03 $ & $  0.27 \pm 0.04 $ & Neutral\tablenotemark{r} & 22\% Water\tablenotemark{4,6,10} \\
 (55565) &                & $ \rm 2002\ AW_{197} $ & $ 18.64 \pm 0.04 $ & $  0.09 \pm 0.05 $ & $  0.18 \pm 0.06 $ & Neutral & Neutral\tablenotemark{1,3} \\
 (55636) &                & $ \rm 2002\ TX_{300} $ & $ 18.89 \pm 0.05 $ & $ -1.38 \pm 0.22 $ & $ -0.53 \pm 0.19 $ &   Water\tablenotemark{f} & 64\% Water\tablenotemark{1} \\
 (55637) &                & $ \rm 2002\ UX_{25} $  & $ 18.48 \pm 0.05 $ & $ -0.02 \pm 0.06 $ & $  0.12 \pm 0.05 $ & Neutral & Neutral\tablenotemark{1} \\
 (55638) &                & $ \rm 2002\ VE_{95} $  & $ 18.22 \pm 0.02 $ & $ -0.08 \pm 0.03 $ & $ -0.06 \pm 0.03 $ & Neutral &  9\% Water\tablenotemark{1} \\
 (65489) &           Ceto & $ \rm 2003\ FX_{128} $ & $ 19.61 \pm 0.06 $ & $  0.09 \pm 0.08 $ & $  0.23 \pm 0.13 $ & Neutral & 14\% Water\tablenotemark{1} \\
 (84522) &                & $ \rm 2002\ TC_{302} $ & $ 18.93 \pm 0.02 $ & $ -0.02 \pm 0.06 $ & $ -0.13 \pm 0.05 $ & Neutral & Neutral\tablenotemark{1} \\
 (84922) &                & $ \rm 2003\ VS_{2} $   & $ 18.19 \pm 0.03 $ & $  0.20 \pm 0.05 $ & $  0.11 \pm 0.08 $ & Neutral &  7\% Water\tablenotemark{1} \\
 (90377) &          Sedna & $ \rm 2003\ VB_{12} $  & $ 19.06 \pm 0.04 $ & $  0.02 \pm 0.05 $ & $ -0.17 \pm 0.09 $ & Neutral & $\sim 10\%$ Water\tablenotemark{7} \\
 (90482) &          Orcus & $ \rm 2004\ DW_{} $    & $ 18.12 \pm 0.03 $ & $ -0.51 \pm 0.04 $ & $ -0.13 \pm 0.05 $ &   Water & $\sim 20\%$ Water\tablenotemark{9} \\
 (90568) &                & $ \rm 2004\ GV_{9} $   & $ 18.55 \pm 0.03 $ & $  0.14 \pm 0.06 $ & $  0.29 \pm 0.07 $ & Neutral & \\
(119951) &                & $ \rm 2002\ KX_{14} $  & $ 19.04 \pm 0.07 $ & $ -0.04 \pm 0.09 $ & $  0.13 \pm 0.10 $ & Neutral & Neutral\tablenotemark{1} \\
(120132) &                & $ \rm 2003\ FY_{128} $ & $ 19.09 \pm 0.03 $ & $  0.04 \pm 0.09 $ & $  0.12 \pm 0.08 $ & Neutral & Neutral\tablenotemark{1,3} \\
(120178) &                & $ \rm 2003\ OP_{32} $  & $ 19.37 \pm 0.02 $ & $ -1.40 \pm 0.28 $ & $ -0.55 \pm 0.14 $ &   Water\tablenotemark{f} & 74\% Water\tablenotemark{1} \\
(120347) &                & $ \rm 2004\ SB_{60} $  & $ 19.53 \pm 0.05 $ & $  0.22 \pm 0.06 $ & $  0.20 \pm 0.07 $ & Neutral & Neutral\tablenotemark{2} \\
(120348) &                & $ \rm 2004\ TY_{364} $ & $ 18.75 \pm 0.03 $ & $ -0.11 \pm 0.06 $ & $ -0.21 \pm 0.07 $ & Neutral & Neutral\tablenotemark{1} \\
(136108) &         Haumea & $ \rm 2003\ EL_{61} $  & $ 16.73 \pm 0.03 $ & $ -0.85 \pm 0.04 $ & $ -0.25 \pm 0.05 $ &   Water\tablenotemark{f} & $\sim 60\%$ Water\tablenotemark{4} \\
(136199) &           Eris & $ \rm 2003\ UB_{313} $ & $ 17.99 \pm 0.03 $ & $ -0.01 \pm 0.05 $ & $ -1.57 \pm 0.12 $ & Methane & $\sim 95\%$ Methane\tablenotemark{8} \\
(136472) &       Makemake & $ \rm 2005\ FY_{9} $   & $ 16.56 \pm 0.02 $ & $ -0.18 \pm 0.04 $ & $ -1.59 \pm 0.07 $ & Methane & $\sim 85\%$ Methane\tablenotemark{5} \\
(145451) &                & $ \rm 2005\ RM_{43} $  & $ 19.08 \pm 0.02 $ & $ -0.22 \pm 0.06 $ & $  0.05 \pm 0.09 $ & Neutral & \\
(145452) &                & $ \rm 2005\ RN_{43} $  & $ 18.43 \pm 0.02 $ & $  0.11 \pm 0.03 $ & $  0.08 \pm 0.03 $ & Neutral & \\
(145453) &                & $ \rm 2005\ RR_{43} $  & $ 19.20 \pm 0.02 $ & $ -1.38 \pm 0.17 $ & $ -0.62 \pm 0.15 $ &   Water\tablenotemark{f} & 65\% Water\tablenotemark{1} \\
(145480) &                & $ \rm 2005\ TB_{190} $ & $ 19.44 \pm 0.04 $ & $  0.00 \pm 0.07 $ & $ -0.08 \pm 0.09 $ & Neutral & \\
(174567) &                & $ \rm 2003\ MW_{12} $  & $ 18.89 \pm 0.02 $ & $  0.08 \pm 0.06 $ & $ -0.08 \pm 0.07 $ & Neutral & \\
(175113) &                & $ \rm 2004\ PF_{115} $ & $ 18.71 \pm 0.02 $ & $  0.02 \pm 0.03 $ & $ -0.04 \pm 0.04 $ & Neutral\tablenotemark{r} & \\
(202421) &                & $ \rm 2005\ UQ_{513} $ & $ 19.04 \pm 0.04 $ & $  0.10 \pm 0.05 $ & $  0.06 \pm 0.06 $ & Neutral\tablenotemark{r} & \\
(208996) &                & $ \rm 2003\ AZ_{84} $  & $ 19.37 \pm 0.03 $ & $ -0.11 \pm 0.08 $ & $  0.08 \pm 0.11 $ & Neutral & 18\%,$\sim 25\%$ Water\tablenotemark{1,3} \\
(225088) &                & $ \rm 2007\ OR_{10} $  & $ 19.06 \pm 0.03 $ & $ -0.15 \pm 0.05 $ & $ -0.08 \pm 0.06 $ & Neutral & \\
\enddata
\tablenotetext{a}{Reported absorption refers to the percentage depth
  of the absorption feature, since surface fraction is a highly
  model-dependent quantity.}
\tablenotetext{f}{{Likely} Haumea family member based on composition
  (this work) and dynamics \citep{2007AJ....134.2160R}.}
\tablenotetext{r}{We find these three objects to be extremely red in
  the visible, indicating that a continuum measurement is difficult to
  obtain.  These objects may have more absorption than indicated from
  the $J-H_{2}O$ or $J-CH_{4}$ metrics.  We flag objects as ``red'' if
  they have $Y-J > 0.1$ to at least 3$\sigma$ significance as
  determined from our photometry.  See Section~\ref{ultrared} for
  details.}
\tablecomments{This table lists all objects observed and their surface
  types based upon our custom photometric system (this work).
  Spectroscopic surface types are based on other works as referenced
  in the table.  Haumea family members are identified as being both
  water-ice rich and having orbital parameters similar to Haumea.
  References --- (1) \cite{2008AJ....135...55B}, (2)
  \cite{2008ApJ...684L.107S}, (3) \cite{2010Icar..208..945M}, (4)
  \cite{2007ApJ...655.1172T}, (5) \cite{2007AJ....133..284B}, (6)
  \cite{2009AA...501..349D}, (7) \cite{2007DPS....39.4906T}, (8)
  \cite{2005ApJ...635L..97B}, (9) \cite{2005ApJ...627.1057T}, (10)
  \cite{2004Natur.432..731J}, (11) \cite{2001AJ....122.2099J}, (12)
  \cite{2007Natur.446..294B}, (13) \cite{2005AA...437.1115D}, (14)
  \cite{2000AJ....119..970N}}
\label{resultstable}
\end{deluxetable}

\clearpage
\begin{deluxetable}{lllllll}
\tabletypesize{\footnotesize}
\tablecolumns{7}
\tablewidth{6in}
\tablecaption{Ice Detections From This Work}
\tablehead{
\colhead{Object} & \colhead{Object} & \colhead{Provisional} & \colhead{$a$\tablenotemark{*}} & \colhead{$e$\tablenotemark{*}} & \colhead{$i$\tablenotemark{*}}     & \colhead{Dynamical} \\
\colhead{Number} & \colhead{Name}   & \colhead{Designation} & \colhead{[AU]}                  & \colhead{}                     & \colhead{[\degr]}                  & \colhead{Class}}
\startdata
\hline \hline
Haumea family \\
 \hline
         &                & $ \rm 2003\ UZ_{117} $ & 44.26 & 0.13 & 27.88 & Haumea \\
         &                & $ \rm 2005\ CB_{79} $  & 43.27 & 0.13 & 27.17 & Haumea \\
 (19308) &                & $ \rm 1996\ TO_{66} $  & 43.32 & 0.12 & 28.02 & Haumea \\
 (24835) &                & $ \rm 1995\ SM_{55} $  & 41.84 & 0.10 & 26.85 & Haumea \\
 (55636) &                & $ \rm 2002\ TX_{300} $ & 43.29 & 0.13 & 26.98 & Haumea \\
(120178) &                & $ \rm 2003\ OP_{32} $  & 43.24 & 0.10 & 27.05 & Haumea \\
(136108) &         Haumea & $ \rm 2003\ EL_{61} $  & 43.10 & 0.19 & 26.85 & Haumea \\
(145453) &                & $ \rm 2005\ RR_{43} $  & 43.27 & 0.13 & 27.07 & Haumea \\
         &                & $ \rm 2009\ YE_{7} $   & 44.43 & 0.13 & 27.99 & Haumea\\
\hline \hline
Non-Haumea Water Ice\\
\hline 
 (90482) &          Orcus & $ \rm 2004\ DW_{} $    & 39.17 & 0.23 & 20.56 & 3:2 Resonance \\
 (28978) &          Ixion & $ \rm 2001\ KX_{76} $  & 39.62 & 0.25 & 19.65 & 3:2 Resonance \\
\hline \hline
Methane objects \\
 \hline 
(136199) &           Eris & $ \rm 2003\ UB_{313} $ & 68.01 & 0.43 & 43.83 & Scattered \\
(136472) &       Makemake & $ \rm 2005\ FY_{9} $   & 45.36 & 0.16 & 29.01 & Classical \\
\enddata

\label{familymembers}

\tablenotetext{*}{For the Haumea family members except for $\rm
  2009\ YE_{7}$, $a$, $e$ and $i$ are the proper orbital elements
  semimajor axis, eccentricity and inclination, respectively, as
  presented by \cite{2007AJ....134.2160R}.  For $\rm 2009\ YE_{7}$,
  proper orbital elements are presented for the first time in this
  work, and are an average of 20 million years of orbital motion
  computed by the MERCURY symplectic integrator
  \citep{1999MNRAS.304..793C}.  The orbital elements for the
  non-Haumea family members are osculating elements provided by the
  Minor Planet Center.}

\tablecomments{Bodies from our survey with significant amount of ices
  on their surfaces.  The top section of the table are all Haumea
  family members based on their (1) water ice absorption and (2) the
  proximity of their proper orbital elements to Haumea as identified
  by \cite{2007AJ....134.2160R}, with the exception of $\rm
  2009\ YE_{7}$ which is identified in this work based on the proper
  orbital elements we computed, the water ice absorption depth and the
  optical colors of the body.  Orcus and Ixion are two non-family
  members with significant water ice absorption.}
\end{deluxetable}

\clearpage
\begin{deluxetable}{llllll}
\tabletypesize{\footnotesize}
\tablecolumns{6}
\tablewidth{4.5in}
\tablecaption{Near-infrared Observations of KBOs Near the Haumea Family}
\tablehead{
\colhead{Object} & \colhead{Object} & \colhead{Provisional} & \colhead{$\Delta \nu_{\rm min}$\tablenotemark{*}} & \colhead{$T_p$}\tablenotemark{*} & \colhead{Water} \\
\colhead{Number} & \colhead{Name}   & \colhead{Designation} & \colhead{[$\rm \mbox{m s}^{-1}$]}                & \colhead{}                       & \colhead{Ice}}
\startdata
(136108) &         Haumea & $ \rm 2003\ EL_{61} $  &     0 & 2.83 & Yes \\
 (19308) &                & $ \rm 1996\ TO_{66} $  &  15.0 & 2.83 & Yes \\
(145453) &                & $ \rm 2005\ RR_{43} $  &  58.0 & 2.84 & Yes \\
         &                & $ \rm 2003\ UZ_{117} $ &  60.8 & 2.84 & Yes \\
         &                & $ \rm 2005\ CB_{79} $  &  66.5 & 2.84 & Yes \\
 (55636) &                & $ \rm 2002\ TX_{300} $ &  68.4 & 2.84 & Yes \\
(120178) &                & $ \rm 2003\ OP_{32} $  &  91.4 & 2.85 & Yes \\
         &                & $ \rm 2009\ YE_{7} $   & $\sim 100$ & 2.84 & Yes \\
 (24835) &                & $ \rm 1995\ SM_{55} $  & 123.3 & 2.84 & Yes \\
\hline
(202421) &                & $ \rm 2005\ UQ_{513} $ &  39.0 & 2.84 & No \\
(136472) &       Makemake & $ \rm 2005\ FY_{9} $   & 118.0 & 2.84 & No \\
         &                & $ \rm 2004\ PT_{107} $ & 161.9 & 2.83 & No \\
(120347) &                & $ \rm 2004\ SB_{60} $  & 218.5 & 2.86 & No \\
 (40314) &                & $ \rm 1999\ KR_{16} $  & 242.9 & 2.84 & No \\
\enddata
\label{haumeaobs}

\tablenotetext{*}{As computed in \cite{2007AJ....134.2160R}, excepting
  $\rm 2009\ YE_{7}$.  The explicit computation of $\Delta \nu_{\rm
    min}$ for $\rm 2009\ YE_{7}$ is beyond the scope of this work.
  However we include a crude estimate based on the fact that the
  proper orbital elements of $\rm 2009\ YE_{7}$ are close to those of
  $\rm 2003\ UZ_{117}$ and well within the $\Delta \nu = 150 \mbox{ m
    s}^{-1}$ region of Figure 1 from \cite{2007AJ....134.2160R}.  We
  find in this work that $\rm 2009\ YE_{7}$ is a Haumea family
  member.}

\tablecomments{We list the bodies we observed that were also
  identified in \cite{2007AJ....134.2160R} as potential family members
  based on their dynamical association with Haumea, ordered by $\Delta
  \nu_{\rm min}$.  The bodies with water ice detected appear to have
  lower $\Delta \nu_{\rm min}$ compared to those that have no water
  ice, although this result is not significant at the $3 \sigma$
  level.}
\end{deluxetable}

\clearpage
\begin{deluxetable}{lr}
\tablecolumns{2}
\tablewidth{4in}
\tablecaption{Visible Photometry for $\rm 2009\ YE_{7}$ From This Work}
\tablehead{
\colhead{Bandpass} & \colhead{Magnitude}}
\startdata
$r'$    & $21.55 \pm 0.03$ \\
$g'-r'$ &  $0.45 \pm 0.04$ \\
$r'-i'$ &  $0.30 \pm 0.05$ \\
$R$     & $21.35 \pm 0.03$\tablenotemark{*} \\
$B-R$   &  $1.03 \pm 0.04$\tablenotemark{*} \\
$V-R$   &  $0.38 \pm 0.04$\tablenotemark{*} \\
$R-I$   &  $0.51 \pm 0.05$\tablenotemark{*} \\
$m_R(1,1,0\degr)$ & $4.13 \pm 0.05$\tablenotemark{*} \\
\enddata
\label{ye7visible}
\tablenotetext{*}{Sloan colors were converted to the
  Johnson-Morgan-Cousins BVRI color system using transfer equations
  from \cite{2002AJ....123.2121S}, as described in
  \cite{2010AJ....139.1394S}.}
\tablecomments{At the time of the observations, $\rm 2009\ YE_{7}$ was
  at heliocentric distance 50.665 AU, geocentric distance 51.153 AU,
  phase angle 0.979 degrees.  $m_{R}(1,1,0\degr)$ refers to a
  hypothetical red magnitude at 1AU heliocentric and geocentric
  distance and phase angle of $0\degr$.}
\end{deluxetable}

\clearpage

\clearpage

\clearpage

\clearpage

\clearpage

\clearpage

\clearpage

\end{document}